\documentclass[aps,amsmath,amssymb,amsfonts,tightenlines,prd,10pt,preprint,superscriptaddress,nofootinbib]{revtex4-1}

\usepackage{bm, slashed}
\usepackage{graphicx}
\usepackage{epic,eepic}
\usepackage{enumitem}

\newcounter{incre}
\newcommand{\benum}{\begin{enumerate}[label=\roman{enumi}),ref=\roman{enumi}] \setcounter{enumi}{\value{incre}}}
\newcommand{\eenum}{\setcounter{incre}{\value{enumi}}\end{enumerate}}
\newcommand{\beq}{\begin{equation}}
\newcommand{\eeq}{\end{equation}}
\newcommand{\ArcTan}{\tan^{-1}}
\newcommand{\bea}{\begin{eqnarray}}
\newcommand{\eea}{\end{eqnarray}}

\newcommand{\bGamma}{\bar{\Gamma}}

\def\coradd{Laboratory for Elementary-Particle Physics, Cornell University, Ithaca, N.Y.}

\begin{document}

\title{Kinematic Edges with Flavor Oscillation and Non-Zero Widths}
\author{Yuval Grossman}
\email{yg73@cornell.edu}
\affiliation{\coradd}
\author{Mario Martone}
\email{mcm293@cornell.edu}
\affiliation{\coradd}
\affiliation{Dipartimento di Scienze Fisiche, University of Napoli and INFN,
Via Cinthia I-80126, Napoli}
\affiliation{Department of Physics,
Syracuse University, Syracuse, N.Y.\vspace*{6mm}}
\author{Dean J. Robinson}
\email{djr233@cornell.edu}
\affiliation{\coradd}
\date{\today}

\begin{abstract}
\vspace*{6mm}

Kinematic edges in cascade decays provide a probe of the masses of new
particles. In some new physics scenarios the decay chain involves
intermediate particles of different flavors that can mix and oscillate. We discuss
the implication of such oscillation, and in particular its interplay with
the non-zero widths of the particles. We derive explicit formulae for
differential decay rates involving both non-zero widths and oscillation,
and show that in the case where the mass difference between the
intermediate particle is of the order of their widths, both
oscillation and width effects are important. An examination of the
physical observables contained in these differential decay rates is
provided. We calculate differential decay rates for cases in
which the intermediate particles are either scalars or fermions.

\end{abstract}

\maketitle

\section{Introduction}
Many new physics scenarios predict cascade decays of new heavy degrees of freedom into Standard Model (SM) particles. A canonical example is the cascade decay of a squark into a quark plus two leptons and a neutralino $\tilde{q} \to q\tilde{\chi}_2^0 \to ql\tilde l \to q l l \tilde{\chi}_1^0$. It is well-known that for a cascade decay of the general form 
\begin{equation}
	\label{eqn:CD}
	A \to XB \to XYC~,
\end{equation}
where $X$ and $Y$ are massless SM particles and $m_A>m_B>m_C$, the
differential decay rate, $d\Gamma_A/ds$, possesses a kinematic edge
located at
\cite{Allanach:2000,Hinchliffe:1997,Meade:2006,Gjelsten:2007,Burns:2009,Lester:2007}
\begin{equation}
	\label{eqn:KE}
	s= \frac{(m_A^2 - m_B^2)(m_B^2 - m_C^2)}{m_B^2},\qquad s\equiv(p_X + p_Y)^2~.
\end{equation}
This kinematic edge is in essence a step function in the differential
decay rate distribution, and it arises due to kinematic upper bounds on the on-shellness of the intermediate exchanged particle,
$B$. The location of the kinematic edge provides an indirect means to
either measure or constrain the masses of the $A$, $B$ and $C$
particles involved in the cascade. This mass measurement technique is
called kinematic edge or endpoint method
\cite{Weiglein:2006,Barr:2010,Gjelsten:2004,Gjelsten:2005,Gjelsten:2006,Lester:2006,Paige:1996}. It
is particularly important in the case that $C$ is invisible, in which case the particle masses cannot
be measured directly.

In order to derive the kinematic edge in Eq. (\ref{eqn:KE}), one must
assume that $B$ is an on-shell mass eigenstate. This is a natural and
plausible assumption to make, but it neglects the fact that $B$ must
also have a non-zero width, $\Gamma_B>0$. One expects a non-zero width
for $B$ to smear the kinematic edge, because such a width smears out
the invariant mass range within which $B$ can be on-shell. However,
for all phenomenologically important scenarios $\Gamma_B \ll m_B$, so this smearing effect is considered to be small, and for this reason the role
of $\Gamma_B$ has been usually neglected. In some previous analyses,
$\Gamma_B$ has been incorporated into the differential decay rate by
convolving the kinematic edge with a Breit-Wigner
distribution \cite{Miller:2006}.

In many well-motivated theories the field $B$ has not one but several
flavors, which means that $B$ is a superposition of multiple mass
eigenstates that may mix together and oscillate. For example, this
scenario is predicted in various SUSY theories
\cite{ArkaniHamed:1996au,ArkaniHamed:1997km,Hisano:2002iy,Feng:2007ke,Hiller:2008sv},
and many proposals of ways to measure mass splittings, mixing and oscillation have
been presented previously
\cite{Agashe:1999bm,Hinchliffe:2000np,Kitano:2008en,Feng:2009yq,Feng:2009bd}. For
just two flavors, it is well-known that the importance of the
interference --- the oscillation --- between the mass eigenstates, denoted
$B_1$ and $B_2$, is characterized by the dimensionless parameter
\begin{equation}
	x \equiv \frac{\Delta m}{\bGamma}~,
\end{equation}
that is the ratio of the $B_{1,2}$ mass splitting to their average
decay rate. In the case that $x \ll1$ or $x\gg1$, oscillation is
respectively unimportant because the oscillation length scale is too
long or the oscillation is washed out. Oscillation effects, however, are
significant in the case that $x \sim 1$. Due to the dependence of $x$
on both the mass splitting and the decay rates, non-zero width and
flavor oscillation effects cannot be independently considered. In other
words, analysis of flavor oscillation requires the incorporation of the
non-zero $B$ widths into the computation of the differential decay
rates. 

If the flavors do not oscillate significantly, or if interference
is negligible due to $x\gg1$, then we simply expect $d\Gamma_A/ds$ to feature
multiple, distinct kinematic edges, each corresponding to a single
mass eigenstate, and the role of the widths should be
unimportant. A detailed analysis of the physical information contained in the differential decay rates for the limit $x \to \infty$, in which oscillation and widths are negligible, has been conducted in Ref. \cite{Galon:2011wh}. However, if oscillation is significant, then we not only
expect interference terms to become important, but we also expect the
form of $d\Gamma_A/ds$ to be smeared by the non-negligible widths.

The purpose of this paper is to examine this prediction and its consequences in detail. We do this for the case of two-flavor mixing, with $B_{1,2}$ either scalars or spin-1/2 fermions that interact with the external particles $A$, $C$, $X$, and $Y$ via Yukawa-type interactions. In Sec. \ref{sec:FW} we first redevelop the kinematic edge formalism via a field theoretic approach, accounting for the finite width with a Breit-Wigner propagator for $B$. For the simple case of a scalar $\phi^3$ interaction with a single intermediate $B$, we show explicitly how the kinematic edge arises in the $\Gamma_B \to 0$ limit. We also show that the kinematic edges have their own well-defined `edge width' that is a function of $\Gamma_B/m_B$, permitting us to quantify how much the kinematic edge is smeared for a given $B$ width.

In Sec. \ref{sec:FM}  we introduce two-flavor mixing for both the scalar and fermion cases, and present the corresponding explicit results for $d\Gamma_A/ds$ in detail. As expected, we find that interference between the mass eigenstates is important only for the regime $x \sim 1$. We also rederive the result that in the fermion case, spin correlations between $X$ and $Y$ alter the shape of $d\Gamma_A/ds$ dramatically compared to the scalar $B$ case, and verify that in the case of a vectorial coupling, the two-flavor fermionic case corresponds to the scalar one (see e.g. \cite{Wang:2008sw}). We present the explicit and detailed derivations of the results presented in Sec. \ref{sec:FM} in Appendix \ref{sec:AppA}.

Finally, in Sec. \ref{sec:OSD} we briefly explore the physical
observables contained in the differential decay rates with oscillation, and
how these may be used to constrain or measure the oscillation parameters,
the masses, and the decay rates. We also propose a kinematic edge
resolution criterion, which specifies under what conditions the two
kinematic edges can be distinguished. We show that in most regions of parameter space the edge resolution criterion is simply
$x >1$, which aligns with our expectation that the edges should be resolvable when interference is negligible. We consider as an example in Sec. \ref{sec:OSD} the
special case that $B$ consists of two flavors of sleptons, with oscillation
parameters as motivated by gauge mediation SUSY breaking
theories. Along with the kinematic edge constraints, we show that: the
degree of oscillation --- the magnitude of $x$ --- can be determined directly
from the widths of the edges; the ratio of the kinematic edge step heights provides three
observables, which uniquely constrain a parameter subspace involving
$x$, the mass splitting and the mixing angle; and that the $s=0$
intercepts provide two other physical observables, which are
measurable even if the edges cannot be resolved. Lastly, we show that
for the alternative case that the $B$s are fermions, the parameter
space is constrained by these observables in an identical fashion to
the scalars, the only difference being that the fermion parameter
space is enlarged by one extra dimension compared to the scalar case.

\section{Non-Zero Width}
\label{sec:FW}
In this section, we examine the role of a non-zero width $\Gamma_B$, and how it affects the sharp kinematic edge of Eq. (\ref{eqn:KE}). The usual derivation of Eq. (\ref{eqn:KE}) requires three assumptions: $X$ and $Y$ are massless; energy-momentum conservation; and, crucially, that $B$ is on-shell. One may then derive Eq. (\ref{eqn:KE}) from kinematics alone. To include the finite width, we must instead perform an explicit field-theoretic computation of the differential decay rate $d\Gamma_A/ds$. 

\begin{figure}[t]
\includegraphics[scale=0.8]{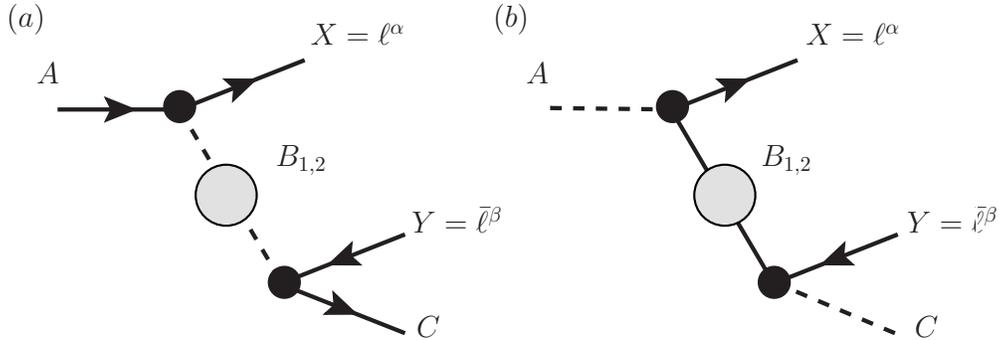}
\caption{Decay amplitudes for the particle $A$ in the case that the intermediate states $B_{1,2}$ are either (a) scalars or (b) fermions.}
\label{fig:SFD}
\end{figure}

\subsection{Breit-Wigner Approximation}
In this paper, we will be mainly concerned with the differential decay rates associated with the two amplitudes shown in Fig. \ref{fig:SFD}, in which the intermediate exchanged particles $B_{1,2}$ are respectively either scalars or fermions, and have Yukawa-type couplings to the external fields $A$, $X$, $Y$ and $C$.  We assume that $X$ and $Y$ are both leptons, which are denoted by the fields $\ell^\alpha$. We also assume, unless stated otherwise, that $X$ and $Y$ are massless
\begin{equation}
	\label{eqn:MXY}
	m_X = m_Y = 0~,
\end{equation}
while the particles $A$, $B_{1,2}$ and $C$ have cascade-ordered masses
\begin{equation}
	\label{eqn:COM}
	m_A > m_{1,2} > m_C~.
\end{equation}
To simplify notation we drop henceforth the subscript $A$ from the decay rate, so that $\Gamma_A\equiv\Gamma$. 

Let us consider the propagator for $B$ in the field-theoretic approach. Throughout the paper we shall always assume that
\begin{equation}
	\label{eqn:SWA}
	\Gamma_B/m_B \ll 1~.
\end{equation}
It is well-known that in the small width regime (\ref{eqn:SWA}), and provided $p_B^2 \simeq m_B^2$, the exact propagator for an unstable particle is well-approximated by the Breit-Wigner propagator. (In a formal way, this approximation amounts to the leading order term of the propagator's Laurent series expanded around the pole $p^2=m^2-im\Gamma$.) In the same spirit as the saddle point approximation, integrals of amplitudes involving the propagator --- precisely the objects with which we are concerned in this paper --- are well-approximated by the Breit-Wigner function, \emph{provided} the domain of phase space integration includes the Breit-Wigner maximum $p_B^2 = m_B^2$. The reason is that the dominant contribution to such integrals is then due to an interval containing $p_B^2 = m_B^2$, and the Breit-Wigner is the leading order approximation within this interval. Outside this interval, the contribution to the integral by both the propagator and the Breit-Wigner are negligible, so we can approximate the propagator by the Breit-Wigner for all $p_B^2$, even though the Breit-Wigner tails may be poor approximations to the exact propagator tails.  This argument extends naturally to objects involving sums of different propagators, that is, the differential decay rate for multiple flavors.  

Inclusion of the Breit-Wigner maximum  $p_B^2 = m_B^2$ within the domain of integration is guaranteed by assumption (\ref{eqn:COM}). Hence for any number of flavors, finite width effects in the differential decay rate are approximated to leading order in $\Gamma_B/m_B$ by use of the Breit-Wigner propagators. 

For just two flavors, note that this result holds for all values of $x$. In particular, it is well-known that if the maxima are well-separated such that only the propagator tails overlap --- i.e. $x$ is large --- then the Breit-Wigner approximation is a poor approximation to the corresponding interference term.  However, we emphasize that if only the tails of the propagators overlap, then the corresponding interference term is also always negligible compared to the other terms contributing to the differential decay rate. As a result the failure of the Breit-Wigner approximation in the interference term for large $x$ results in a negligible correction to the overall differential decay rate. So the Breit-Wigner approximation is a valid leading order approximation for all $x$.

For the scalar, the Breit-Wigner approximation is
\begin{equation}
	\label{eqn:BWS}
	D_{\textrm{sc}}(p^2) = \frac{i}{p^2 - m^2 + im\Gamma}~,
\end{equation}
while for a fermion 
\begin{equation}
	\label{eqn:FBW}
	 D_{\textrm{f}}(\slashed{p}) 
	 = i\frac{\slashed{p} + \sqrt{m^2 - im\Gamma}}{p^2 - m^2 + im\Gamma}
	  \simeq i\frac{\slashed{p} + m -i\Gamma/2}{p^2 - m^2 + im\Gamma}~,
\end{equation}
due to Eq. (\ref{eqn:SWA}). Note that we keep the $i\Gamma/2$ term as it is not necessarily small compared to $\slashed{p} + m$, and in fact it produces leading order contributions in the case of intermediate particle oscillation.

\subsection{$\phi^3$ Interaction}
\label{sec:P3}

To study the smearing of the kinematic edge due to $\Gamma_B \not= 0$, it is instructive to first consider a toy $\phi^3$ interaction calculation in which all the particles involved in the cascade are scalars and computational technicalities are therefore simplified. For further simplicity, we assume $B$ is just a single scalar mass eigenstate. 

Consider such a $\phi^3$ scalar interaction, of the form 
\begin{equation}
\mathcal{L}=g_X\phi_A\phi_B\phi_X + g_Y\phi_C\phi_B\phi_Y~.
\end{equation} 
The amplitude for the decay of $A$ is
\begin{equation}
	i\mathcal{M} = \frac{ig_Xg_Y}{p_B^2 - m_B^2 + im_B\Gamma_B}~.
\end{equation}
Squaring this amplitude and integrating over phase space, one finds
\begin{equation}
	\label{eqn:DDRPC}
	\frac{d\Gamma}{ds} = \frac{g_X^2g_Y^2}{32(2\pi)^3m_A^3}\frac{1}{m_B\Gamma_B}\tan^{-1}\bigg(\frac{m_B}{\Gamma_B}\eta\bigg)\Bigg|_{\eta = \eta_-(s)}^{\eta = \eta_+(s)}~,
\end{equation}
where
\begin{align}
	\eta_{\pm}(s) & \equiv 1+ \xi(s) \pm \sqrt{\xi(s)^2 - m_A^2m_C^2/m_B^4}\notag\\
	\xi(s) & \equiv \frac{s - (m_A^2 + m_C^2)}{2m_B^2}~. \label{eqn:DEX}
\end{align}
We report the details of this calculation in Appendix \ref{sec:AppA}.

It is important to note that $\Gamma_B \propto g_Y^2/m_B$ ($g_{X,Y}$ have mass dimension one here), so in the limit $\Gamma_B/m_B \to 0$, $g_Y^2/m_B\Gamma_B$ is finite.  This means that Eq. (\ref{eqn:DDRPC}) has a well-defined normalization in the $\Gamma_B/m_B \to 0$ limit. To encode this explicitly, we therefore write 
\begin{equation}
	\label{eqn:PCN}
	\frac{g_Y^2}{m_B\Gamma_B} = \tilde{g}_Y^2~,
\end{equation}
which is dimensionless. Adopting this convention, Eq. (\ref{eqn:DDRPC}) becomes
\begin{equation}
	\label{eqn:DDRPC2}
	\frac{d\Gamma}{ds} = \frac{g_X^2\tilde{g}_Y^2}{32(2\pi)^3m_A^3}\tan^{-1}\bigg(\frac{m_B}{\Gamma_B}\eta\bigg)\Bigg|_{\eta = \eta_-(s)}^{\eta = \eta_+(s)}~,
\end{equation}
which is manifestly finite in the limit $\Gamma_B/m_B \to 0$. We will employ similar redefinitions of the coupling at the $Y$ vertex throughout the paper.

Overall momentum conservation in the $A$ rest frame requires that $s = m^2_A + m_C^2 - 2E_Cm_A^2$, so it must always be that 
\begin{equation}
	s \le \big(m_A - m_C\big)^2\equiv s_{\textrm{max}}~.
\end{equation}
Observe that $\eta_-(s_{\textrm{max}}) = \eta_+(s_{\textrm{max}})$, and $\eta_{\pm}(s)$ become complex for $s> s_{\textrm{max}}$, so the differential decay rate is always precisely zero for $s \ge s_{\textrm{max}}$ as expected. We emphasize that this maximum is not related to the kinematic edge, but is a distinct kinematic constraint.

Let us now extract the kinematic edge from the differential decay rate. Expanding Eq. (\ref{eqn:DDRPC2}) in $\Gamma_B/m_B \ll 1$ we have
\begin{equation}
	\label{eqn:PCGE}
	\frac{d\Gamma}{ds} \propto \frac{\pi}{2} \big\{\mbox{sgn}\big[\eta_+(s)\big] - \mbox{sgn}\big[\eta_-(s)\big]\big\} - \bigg(\frac{1}{\eta_+(s)} - \frac{1}{\eta_-(s)}\bigg)\frac{\Gamma_B}{m_B} + \ldots~.
\end{equation}
The leading order term produces a step function in $s$: For $\mbox{sgn}[\eta_+(s)] = -\mbox{sgn}[\eta_-(s)]$ the leading order term is $\pi$, while for $\mbox{sgn}[\eta_+(s)] = \mbox{sgn}[\eta_-(s)]$, the leading order term is zero. The location of this step, $s_0$, must therefore satisfy either
\begin{equation}
	\label{eqn:LKE}
	\eta_{+}(s_0) = 0~,\qquad \mbox{or}\qquad \eta_{-}(s_0) = 0~.
\end{equation}
(The ratio of $m_B$ to the geometric mean of $m_A$ and $m_C$ determines which function is zero in Eqs. (\ref{eqn:LKE}). If $m_Am_C/m_B^2 < 1$ then $\eta_-(s_0) = 0$ and $\eta_+(s_0) >0$; if  $m_Am_C/m_B^2 > 1$ then $\eta_+(s_0) = 0$ and $\eta_-(s_0) <0$; while if $m_Am_C/m_B^2 =1$ then both $\eta_{\pm}(s_0) = 0$.)

\begin{figure}[t]
\includegraphics{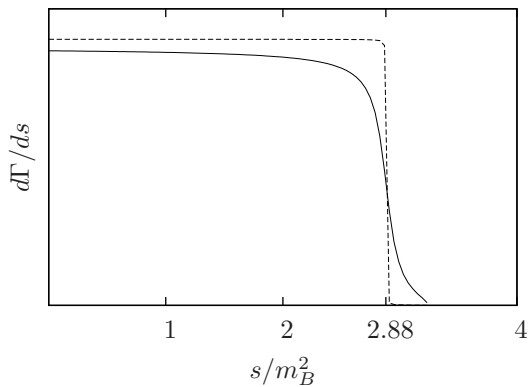}
\caption{Differential decay rate with parameter choice $m_A/m_B = 2$, $m_C/m_B = 0.2$ and $\Gamma_B/m_B = 10^{-1}$ (solid) or $\Gamma_B/m_B = 10^{-3}$ (dashed). The kinematic edge $s_0/m_B^2 = 2.88$ clearly emerges as $\Gamma_B/m_B \to 0$.}
\label{fig:PC}
\end{figure}

Using the definitions (\ref{eqn:DEX}), one may verify that the solution to Eqs. (\ref{eqn:LKE}) is always
\begin{equation}
	\label{eqn:DSZ}
	s_0 = \frac{(m_A^2 - m_B^2)(m_B^2 - m_C^2)}{m_B^2}~,
\end{equation}
which is precisely the expected kinematic edge (\ref{eqn:KE}). We have
therefore shown that the zeroth order contribution in $\Gamma_B/m_B$
to Eq. (\ref{eqn:PCGE}) produces the kinematic edge, while terms of
the order $\Gamma_B/m_B$ and higher smear the edge into
Eq. (\ref{eqn:DDRPC2}). A plot of the differential decay rate for
different $\Gamma_B/m_B$ is shown in
Fig. \ref{fig:PC}.\footnote{Readers expert in the kinematic edge
method may wonder why the differential decay rate in Fig. \ref{fig:PC}
is rectangular in shape, rather than the usual triangle. The reason is
that we have plotted here $d\Gamma/ds$ rather than
$d\Gamma/d\sqrt{s}$, the latter convention being common in the
Literature because the background is usually flat in $\sqrt{s}$. However, the former convention is also used (see e.g. \cite{Wang:2008sw}). Throughout this paper we shall always consider $d\Gamma/ds$.} Notice that if $m_B^2 = m_Am_C$, then $s_0 = s_{\textrm{max}}$, so the kinematic edge collides with the overall kinematic constraint. For all other cases $s_0 < s_{\textrm{max}}$. Note also that the terms of the order $\Gamma_B/m_B$ and higher in Eq. (\ref{eqn:PCGE}) each diverge at the kinematic edge, while the resummed expression in Eq. (\ref{eqn:DDRPC2}) is finite. This occurs because the formal expansion of the Breit-Wigner propagator in powers of $\Gamma_B/m_B$ does not converge. Hence, even though expanding Eq. (\ref{eqn:DDRPC2}) provides us insight into the kinematic edge structure, nonetheless the full expression (\ref{eqn:DDRPC2}) is itself the leading order differential decay rate in $\Gamma_B/m_B$. Similarly, throughout this paper we will not expand the closed form functions produced by the Breit-Wigner approximation, although we will expand their prefactors.

That the kinematic edge arises from the $\Gamma_B/m_B \to 0$ limit of an arctangent function is the main result of this section. We will see below that these functions, and their associated kinematic edges, are a general feature of the differential decay rates considered in this paper. 

\subsection{On-shellness of $B$}
It is also instructive at this point to consider how the on-shellness of $B$ is encoded in this field-theoretic derivation of the kinematic edge. The amplitude squared can be written as
\begin{equation}
	|\mathcal{M}|^2 =\frac{g_X^2g_Y^2}{(p^2 - p_0^2)(p^2 -p_0^{*2})} = \frac{g_X^2g_Y^2}{p_0^2 -p_0^{*2}}\bigg[\frac{1}{p^2 - p_0^2} - \frac{1}{p^2 - p_0^{*2}}\bigg]~,
\end{equation}
where $p_0^2 = m_B^2 -im_B\Gamma_B$ and the $^*$ indicates complex conjugation. The key observation is that if one takes the $\Gamma_B/m_B \to 0$ limit before rather than after computation of $d\Gamma/ds$, one finds that the amplitude squared is
\begin{equation}
	|\mathcal{M}|^2 = \lim_{\Gamma_B \to 0}\frac{g_X^2g_Y^2}{m_B\Gamma_B}\mbox{Im}\bigg[\frac{1}{p^2 - m_B^2 - im_B\Gamma_B}\bigg] =  \frac{2\pi}{m_B^2} g_X^2\tilde{g}_Y^2\delta(p^2 - m_B^2)~.
\end{equation}
We
then see that the differential decay rate in the limit $\Gamma_B/m_B
\to 0$ is simply a phase space integral of a delta function that
forces $B$ to be on-shell. This integral yields a step function as
expected, with the edge occurring at the value of $s$ for which $B$
can no longer be on-shell.

\subsection{Edge Width} 
\label{sec:EW}
Let us now specify how the smearing of the kinematic edge is related to non-zero $\Gamma_B/m_B$. In order to characterize the amount of smearing of the kinematic edge, one may show that the gradient of the differential decay rate, $d^2\Gamma/ds^2$, is of the form
\begin{equation}
	d^2\Gamma/ds^2  = \frac{f(s)}{(s - s_0)^2 + \sigma^2} ~,
\end{equation}
where $f(s)$ is a smooth function of $s$ that is slowly varying compared to the Breit-Wigner factor near the kinematic edge. Hence near the edge
\begin{equation}
	d^2\Gamma/ds^2 \simeq \frac{f(s_0)}{(s - s_0)^2 + \sigma^2}~, 
\end{equation}
in the same spirit as the saddle point approximation. This Breit-Wigner is clearly maximal at the kinematic edge and has full width at half-maximum (FWHM)
\begin{equation}
	\label{eqn:DEW}
	\sigma \simeq 2m_B^2\frac{\Gamma_B}{m_B}\Big|\big(\Gamma_B/m_B\big)^2 + 1 - m_A^2m_C^2/m_B^4\Big|~.
\end{equation}
We call $\sigma$ the edge width. Note that in general $(\Gamma_B/m_B)^2$ is not necessarily small compared to $1 - m_A^2m_C^2/m_B^4$. For the examples shown in Fig. \ref{fig:PC}, we have $\sigma/m_B^2 \simeq 0.2$ or $2 \times 10^{-3}$ respectively, which match the na\"\i vely expected orders of magnitude. The FWHM of $d^2\Gamma/ds^2$ is in principle a measurable quantity: Measurement of $\sigma$ provides constraints on the size of the ratio $\Gamma_B/m_B$, rendering Eq. (\ref{eqn:DEW}) as an important results.

\subsection{Non-zero Width of $A$}
So far in this discussion we have treated $A$ as an on-shell external state. In practice, however, $A$ itself is an intermediate state of a yet larger cascade that started with a heavier mother particle, denoted $A^\prime$. In such a scenario, one measures $d\Gamma_{A^\prime}/ds$ instead of $d\Gamma_A/ds$, and since $A$ has a non-zero width, then the amplitude corresponding to this differential decay rate has the form
\begin{equation}
	\label{eqn:NZWA}
	\bigg(\frac{i}{p_A^2 - m_A^2 +im_A\Gamma_A}\bigg)\bigg(\frac{i}{(p_A-p_X)^2 - m_B^2 +im_B\Gamma_B}\bigg)~.
\end{equation}
Here the smearing due to $\Gamma_A$ convolves with that of $\Gamma_B$: If $\Gamma_A$ is sufficiently large then we expect the $B$ kinematic edge structure to be lost --- i.e. smeared away --- by this convolution. For non-zero $\Gamma_A$, not only does this convolution generally further smear the kinematic edge, but it also means we cannot generally distinguish the smearing effects of $\Gamma_A$ from $\Gamma_B$ in the differential decay rate. That is, the results of Sec. \ref{sec:P3}, which account for smearing due to non-zero $\Gamma_B$ alone, would be invalid. 

Our analysis of non-zero $\Gamma_B$ effects in this section implies that smearing due to $\Gamma_A$ vanishes as the parameter $\Gamma_A/m_A \to 0$. In particular, from the amplitude (\ref{eqn:NZWA}) we expect the $\Gamma_A$ smearing to be negligible if 
\begin{equation}
	\label{eqn:GAGB}
	\Gamma_A/m_A \ll \Gamma_B/m_B~.
\end{equation}
For multiple flavors we require $\Gamma_A/m_A \ll \Gamma_j/m_j$ for all $j$. In this regime, the effect of $A$'s non-zero width on each kinematic edge is negligible compared to the effects of the respective non-zero $B$ widths. 

From Eq. (\ref{eqn:DDRPC2}), we have
\begin{equation}
	\frac{\Gamma_A}{m_A} \sim \frac{g_X^2\tilde{g}_Y^2}{m_A^2}\frac{s_0}{m_A^2} \lesssim \frac{g_X^2\tilde{g}_Y^2}{m_A^2}~,
\end{equation}
since $s_0/m_A^2 < 1$. So smearing due to $A$'s non-zero width is negligible provided the coupling $g_{X}$ is sufficiently small. For the remainder of this paper, we shall always assume Eq. (\ref{eqn:GAGB}) is satisfied for all flavors, so that smearing due to $A$ is negligible.

We shall now use the insight we have gained into finite width effects from this simple $\phi^3$ theory to study the kinematic edge with flavor oscillation. 

\section{Flavor Oscillation}
\label{sec:FM}
In this section we present results for the cases that $B_{1,2}$ are scalars or fermions, which we call the intermediate scalar and intermediate fermion cases respectively.

\subsection{Oscillation Parameters and Small Width Regime}
\label{sec:MP}
Before proceeding, let us define the following usual oscillation parameters in terms of the $B_{1,2}$ mass and decay rates, $m_{1,2}$ and $\Gamma_{1,2}$:
\begin{equation}
	\label{eqn:DM}
	m \equiv \frac{m_2 + m_1}{2}~,\qquad \Delta m \equiv m_2 - m_1~,\qquad\bGamma\equiv\frac{\Gamma_2 + \Gamma_1}{2}~,\qquad  \Delta \Gamma \equiv \Gamma_2 - \Gamma_1~,
\end{equation}
where $m_2 \ge m_1$ and
\begin{equation}
	\label{eqn:DMXYZ}
	x  \equiv \frac{\Delta m}{\bGamma}~,\qquad y  \equiv \frac{\Delta \Gamma}{2\bGamma}~, \qquad z  \equiv \frac{\Delta m}{2m}~.
\end{equation}
Note that $x\ge 0$ is unbounded, while $-1 \le y \le 1$ and $0 \le z
\le 1$. The small width regimes for each mass eigenstate are defined,
as usual, by $\Gamma_j/m_j \ll 1$. It is convenient to define the
parameters
\begin{equation}
	\epsilon_j  \equiv \Gamma_j/m_j~,\qquad \epsilon \equiv \bGamma/m = \frac{m_1\epsilon_1 + m_2\epsilon_2}{m_1 + m_2}~.\notag
\end{equation}
Just as before, we shall always assume small widths $\epsilon_j \ll1$. Observe that since $m_{1,2}$ and $\epsilon_{1,2}$ are positive definite quantities, then this assumption implies $\epsilon \ll 1.$

The four parameters $m$, $x$, $y$ and $z$ are independent so they uniquely specify $m_{1,2}$, $\Gamma_{1,2}$ and $\epsilon_{1,2}$, viz.
\begin{equation}
	\label{eqn:MGXYZ}
	m_{1,2} = m(1\pm z)~,\qquad \Gamma_{1,2} = \frac{2mz}{x}(1\pm y)~,\qquad \epsilon = \frac{2z}{x}~,\qquad \epsilon_{1,2} = \frac{2z}{x}\bigg(\frac{1 \pm y}{1 \pm z}\bigg)~.
\end{equation} 
It is clear that in the small width approximation
\begin{equation}
	\label{eqn:SWCZ}
	z \ll x.
\end{equation} 
Ideally, we may present all the differential decay rates just in terms of the oscillation parameters  $m$, $x$, $y$ and $z$ alone. However, for the sake of compactness and clarity, we shall instead present our results in terms of a mixture of both $m$, $x$, $y$ and $z$ as well as $m_{1,2}$, $\Gamma_{1,2}$, $\epsilon$ and $\epsilon_{1,2}$ with the understanding that the latter may be expressed in terms of the former via Eqs. (\ref{eqn:MGXYZ}). 

Finally, we assume CP conservation. Thus, for two flavors the mixing
matrix $U$, as defined in what follows (cf. Eqs. (\ref{eqn:LSF}) and
(\ref{eqn:LF})), is real orthogonal and has a single physical mixing
angle, $\theta$. We write
\begin{equation}
	\label{eqn:DU}
	U = \bordermatrix{_\alpha\backslash ^i &\mbox{\tiny{1}} &\mbox{\tiny{2}}  \cr \mbox{\tiny{1}} & \cos\theta & \sin\theta \cr  \mbox{\tiny{2}}& -\sin\theta & \cos\theta\cr}~.
\end{equation}

\subsection{Intermediate Scalars}
\label{sec:ISFM}
First consider the differential decay rate due to the amplitude in Fig. \ref{fig:SFD}a. Here $B$ is a superposition of two mass eigenstates with $B_1$ and $B_2$ scalars of two different flavors, while $A$, $X$, $Y$ and $C$ are fermions, with Yukawa-type vertices defined by
\begin{equation}
	\label{eqn:LSF}
	\mathcal{L}^{\textrm{s}} = \bar{\psi}_A\big(g^X_LP_L + g^X_RP_R\big)\ell^\alpha U^{\alpha i *}\phi^i_B + \bar{\psi}_C\big(g^Y_LP_L + g^Y_RP_R\big)\ell^\alpha U^{\alpha i *}	\phi^i_B~. \\
\end{equation}

We will not report here the study of a single intermediate scalar for
this interaction. As shown in Appendix \ref{sec:AppA}, this case
does not differ considerably from the $\phi^3$ case studied in details
in Section \ref{sec:P3}. The only difference is the presence of linear
and logarithmic terms in $s$, which are suppressed by factors of $\epsilon$.

We write the differential decay rate as
\begin{equation}
	\label{eqn:ISDDR}
	\frac{d\Gamma^{\alpha\beta}}{ds}\bigg|_{\textrm{sc}} = \big[(g_L^X)^2 + (g_R^X)^2\big]\big[(\tilde{g}_L^Y)^2 + (\tilde{g}_R^Y)^2\big]\bigg(\frac{d\Gamma^{\alpha\beta}_1}{ds}\bigg|_{\textrm{sc}} + \frac{d\Gamma^{\alpha\beta}_2}{ds}\bigg|_{\textrm{sc}} + \frac{d\Gamma^{\alpha\beta}_{12}}{ds}\bigg|_{\textrm{sc}}\bigg)~.
\end{equation}
The first two terms come respectively from the squared single $B_1$ and $B_2$ contributions. The final term is the interference term for the two different flavors. The subscript `sc' denotes an internal scalar. Note also that just as for the $\phi^3$ case in Eq. (\ref{eqn:PCN}), the combination $(g_{L,R}^Y)^2m/\bGamma$ is finite in the zero width limit (the couplings are now dimensionless), so we write,
\begin{equation}
	\label{eqn:DTG}
	\big(g_{L,R}^Y\big)^2\frac{m}{\bGamma}=\big(\tilde{g}_{L,R}^Y\big)^2~,
\end{equation}
in which $\tilde{g}_{L,R}^Y$ are finite. In Eq. (\ref{eqn:ISDDR}) we have already removed the $1/\epsilon = m/\bGamma$ factor absorbed by $g^Y_{L,R}$ and replaced it with $\tilde{g}^Y_{L,R}$.

To leading order in $\epsilon$ the square terms are
\begin{equation}
	\frac{d\Gamma^{\alpha\beta}_j}{ds}\bigg|_{\textrm{sc}} = \frac{|U^{\alpha j}|^2|U^{\beta j}|^2}{(2\pi)^3}\frac{s_0^j}{m_A^3}\bigg(\frac{1 + (-)^jz}{1 + (-)^jy}\bigg)\tan^{-1}\bigg[\frac{\eta}{\epsilon_j}\bigg]\Bigg|_{\eta=\eta^j_-}^{\eta = \eta^j_+}~,\label{eqn:STIS}
\end{equation}
in which we have defined (cf. Eqs. (\ref{eqn:DEX}) and (\ref{eqn:DSZ}))
\begin{align}
	s_0^j & \equiv  \frac{(m_A^2 - m_j^2)(m_j^2 - m_C^2)}{m_j^2}~,\notag\\
	\xi_j(s) & \equiv \frac{s - (m_A^2 + m_C^2)}{2m_j^2}~,\notag\\
	\eta_{\pm}^j & \equiv 1 + \xi_j(s) \pm \sqrt{\xi_j^2(s) - m_A^2m_C^2/m_j^4}~.\label{eqn:ISIFDSEX}
\end{align}
In Eq. (\ref{eqn:STIS}) we have discarded terms whose coefficients are subleading order in $\epsilon$ or $\epsilon_j$, but just as for the $\phi^3$ example we have not expanded the arctangent function itself, in order to avoid creating artificial divergences at the kinematic edges. 

The interference term is
\begin{equation}
	\label{eqn:DDRIS}
	\frac{d\Gamma^{\alpha\beta}_{12}}{ds}\bigg|_{\textrm{sc}}  = \frac{m^2}{(2\pi)^3m_A^3}\bigg[\frac{U^{\alpha 1}U^{\beta 1}U^{\alpha 2}U^{\beta 2}}{x^2 + 1}\bigg]\sum_{j=1,2}\Bigg\{\mathcal{A}_	{\textrm{sc}}^j\tan^{-1}\bigg[\frac{\eta}{\epsilon_j}\bigg] +\mathcal{B}_{\textrm{sc}}^j\log\Big[\eta^2 +\epsilon_j^2\Big]\Bigg\}\Bigg|_{\eta=\eta^j_-}^{\eta = \eta^j_+}~.
\end{equation}
with coefficients that are to leading order given by
\begin{align}
	\mathcal{A}_{\textrm{sc}}^{j} & =  \frac{(m_A^2 - m^2)(m^2 - m_C^2)}{m^4}~,\notag\\	
	\mathcal{B}_{\textrm{sc}}^{j} &= -(-1)^j\frac{x}{2}\mathcal{A}_{\textrm{sc}}^j~.\label{eqn:ITIS}
\end{align}

Several comments are necessary concerning the interference term and
coefficients presented in Eqs. (\ref{eqn:DDRIS}) and
(\ref{eqn:ITIS}). We emphasize first that the coefficients are valid
only to leading order in $\epsilon$. We have also dropped
contributions of $\mathcal{O}(z)$ and higher to the interference
term prefactors, because their contributions are always suppressed. To
see this, note that the interference terms can be written generally in
the form
\begin{equation}
	\frac{d\Gamma^{\alpha\beta}_{12}}{ds}\bigg|_{\textrm{sc}} \sim \frac{a + bz}{x^2 + 1}~,
\end{equation}
where $a$ and $b$ are arbitrary linear combinations of the arctangent and logarithm functions and $a/b \sim 1$. It is clear that the denominator ensures the interference term is relevant only for $x \lesssim 1$: If instead $x \gg 1$, then the denominator suppresses the entire interference term, so the $\mathcal{O}(z)$ terms are certainly unimportant. In the case that $x \lesssim 1$, then $z \ll 1$ by Eq. (\ref{eqn:SWCZ}). Hence the $\mathcal{O}(z)$ terms are always negligible. We emphasize, however, that we have not discarded the $z$ dependence of the arctangent and logarithm arguments, only that of their prefactors. The ability to drop the $\mathcal{O}(z)$ contributions in the interference term prefactors applies similarly to the intermediate fermion case. The full expressions, including $\mathcal{O}(z)$ terms are reported in Appendix \ref{sec:AIS}.

\begin{figure}[t]
\includegraphics{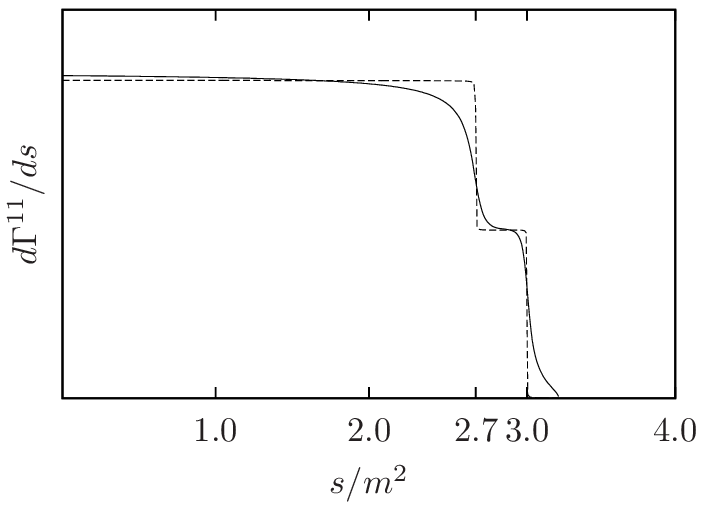}
\includegraphics{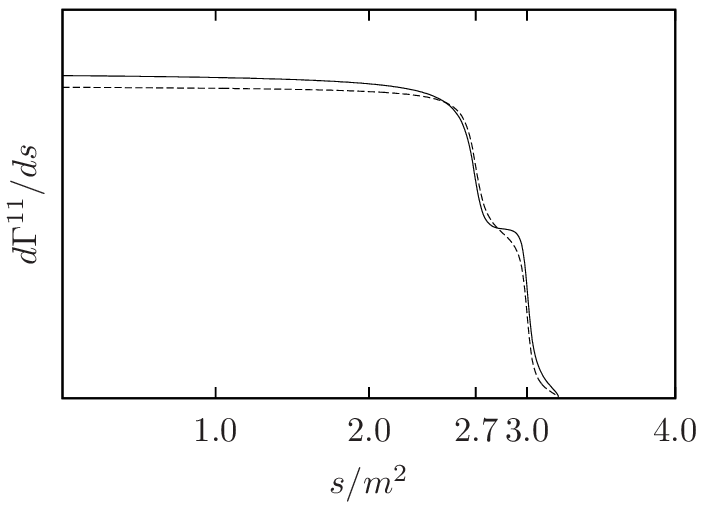}
\caption{Left: Intermediate scalar differential decay rate for $\alpha = \beta = 1$, with the parameter choices $\theta = \pi/4$, $m_A/m = 2$, $m_C/m = 0.2$, $z = y = 0.1$, and $x = 4$ (solid) or $x = 4\times10^2$ (dashed). These correspond to $\epsilon = 5\times10^{-2}$, $5\times 10^{-4}$ respectively. The kinematic edges at $s^1_0/m^2 = 2.7$ and $s_0^2/m^2 = 3.0$ are smeared as $\epsilon$ becomes larger. Right: Intermediate scalar differential decay rate for  the same parameter choices and $x = 4$  with (solid) and without (dashed) the interference terms.}
\label{fig:ISKE}
\end{figure}

The square terms in Eq. (\ref{eqn:STIS}) bear obvious similarities to the $\phi^3$ result in Eq. (\ref{eqn:DDRPC2}). The same analysis applies. Each square term becomes a step function in the zero width limit, producing a kinematic edge at 
\begin{equation}
	\label{eqn:KEFM}
	s = s_0^j \equiv  \frac{(m_A^2 - m_j^2)(m_j^2 - m_C^2)}{m_j^2}~,
\end{equation}
respectively, but these edges are smeared out by non-zero $\epsilon_j$. Eq. (\ref{eqn:MGXYZ}) implies that for a fixed $z$ and $y$, the smaller $x$ is, the larger the $\epsilon_j$ become. Hence in the oscillation regime, $x \sim 1$, the kinematic edges are more smeared compared to the $x\gg1$ case. Similarly, the arctangent contributions to the interference term have edges which are smeared according to the size of the $\epsilon_j$ parameters, with edge widths given by Eq. (\ref{eqn:DEW}). As expected, \emph{strong flavor oscillation smears the kinematic edges.} An example of this smearing of the kinematic edges is shown in Fig. \ref{fig:ISKE} for fixed $z$ but varying $x$. In this figure we have chosen $z =0.1$, a relatively large value, in order that the edges are visually distinct.

As mentioned above, the interference term has a $1/(1+ x^2)$
prefactor, which means that for $x \gg 1$ the entire interference term
is suppressed. Since $x \gg 1$ corresponds to no interference due to flavor
oscillation `wash-out', this is precisely the expected behavior. In
Fig. \ref{fig:ISKE} an example of the relative importance of the
interference term contribution is shown graphically for the case that
$x \sim 1$. As can be seen, for the chosen parameters the interference
terms appear to slightly enhance the sharpness of the
edges. Obviously, as $x$ becomes smaller, the role of the interference
term becomes more significant.

Another new feature of the interference term are the logarithm terms. These logs do not contain kinematic edges. Instead, in terms of $\epsilon$ for fixed $z$, and being careful to include the $1/(x^2 + 1)$ and $\mathcal{B}_{\textrm{sc}} \sim x$ prefactors, the log terms are of form
\begin{equation}
	\frac{d\Gamma^{\alpha\beta}_{12}}{ds}\bigg|_{\textrm{sc,log}} \sim \frac{z\epsilon}{\epsilon^2 + z^2}\log\big(\eta^2 + \epsilon^2\big)\Big|_{\eta_-}^{\eta_+}~.
\end{equation}
From Eq. (\ref{eqn:LKE}), at a kinematic edge we have either $\eta_{\pm} = 0$, so the log terms are largest at a kinematic edge. However, in the $\epsilon \to 0$ (or $x \to \infty$) limit these log terms manifestly vanish. Notice also that for $z \to 0$ with $x$ fixed, the large logs at the kinematic edge cancel due to the different signs of $\mathcal{B}_{\textrm{sc}}^j$.

\subsection{Single Intermediate Fermion}
\label{sec:IFFM}
In contrast to the intermediate scalar case, a single intermediate fermion case presents novelties which are worth discussing. In particular, since the $B$ is now a spin-1/2 particle, we expect spin correlations between $X$ and $Y$. These correlations lead to a differential decay rate with triangular or trapezoidal envelope rather than the square envelope found for the intermediate scalar case. 

Consider the following interactions for a single intermediate fermionic $B$,
\begin{equation}
	\mathcal{L}^{\textrm{f}}  = \bar{\psi}_B\big(g^X_LP_L + g^X_RP_R\big)\ell_X\phi^\dagger_A + \bar{\psi}_B\big(g^Y_LP_L + g^Y_RP_R\big)\ell_Y\phi^\dagger_C~.\label{eqn:LF}
\end{equation}
Squaring the amplitude in Fig. \ref{fig:SFD}b, summing (averaging) over final (initial) spins, and integrating over phase space, one finds differential decay rate of the form
\begin{equation}
	\label{eqn:IFDDRT}
	\frac{d\Gamma}{ds}\bigg|_{\textrm{f}} = \big[(g_L^X\tilde{g}_L^Y)^2 + (g_R^X\tilde{g}_R^Y)^2\big]\frac{d\Gamma}{ds}\bigg|_- + \big[(g_L^X\tilde{g}_R^Y)^2 + (g_R^X\tilde{g}_L^Y)^2\big]\frac{d\Gamma}{ds}\bigg|_+~.
\end{equation}
The subscript `f' denotes an internal fermion. The couplings $\tilde{g}_{L,R}^Y$ are defined as in Eqs. (\ref{eqn:PCN}) and (\ref{eqn:DTG}), and the $m_B/\Gamma_B$ factor has already been similarly factored out of the decay rates $d\Gamma/ds|_{\pm}$. The two terms in Eq. (\ref{eqn:IFDDRT}) arise from two purely chiral interactions: i.e. the cases $g_L^X = g_R^Y = 0$ or $g_R^X = g_R^Y = 0$ respectively. We therefore call $d\Gamma/ds|_{\pm}$ the chiral differential decay rates. The chiral differential decay rates turn out to have respectively a positive or negative slope in $s$, whence the subscript (see Appendix \ref{sec:AppA} for details).

One finds to leading order in $\Gamma_B/m_B \ll 1$,
\begin{align} 
\frac{d\Gamma}{ds}\bigg|_-&= \frac{s_0- s}{(2\pi)^3m^3_A}\ArcTan\bigg[\frac{m_B}{\Gamma_B}\eta\bigg]\bigg|_{\eta=\eta_-(s)}^{\eta=\eta_+(s)} \notag\\
\frac{d\Gamma}{ds}\bigg|_+&=\frac{s}{(2\pi)^3m^3_A}\ \ArcTan\bigg[\frac{m_B}{\Gamma_B}\eta\bigg]\bigg|_{\eta=\eta_-(s)}^{\eta=\eta_+(s)}~,\label{eqn:IFNF}
\end{align}
where $\eta_{\pm}$ and $s_0$ are defined in Eqs. (\ref{eqn:DEX}) and (\ref{eqn:DSZ}) respectively.

\begin{figure}[t]
\includegraphics{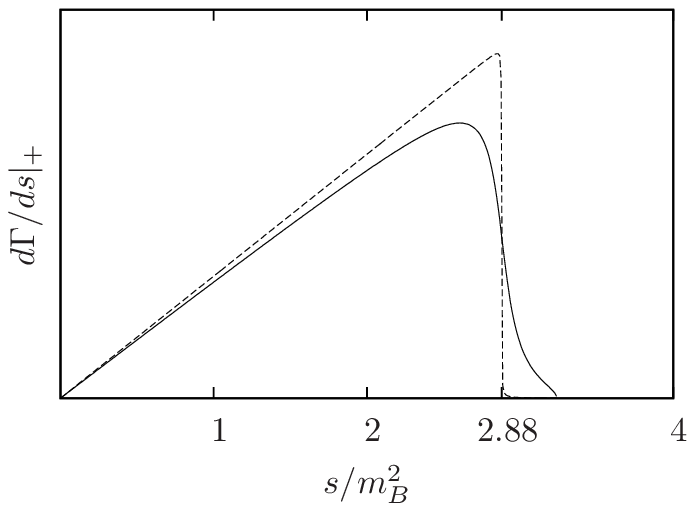}
\includegraphics{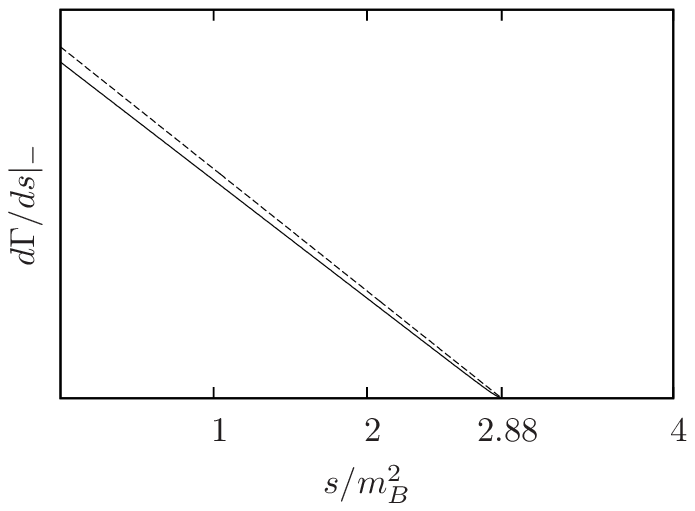}
\caption{Chiral differential decay rates for the cases $g_L^X = g_R^Y = 0$ (left) and $g_R^X = g_L^Y = 0$ (right) with parameter choice $m_A/m_B = 2$, $m_C/m_B = 0.2$ and $\Gamma_B/m_B = 10^{-1}$ (solid) or $\Gamma_B/m_B = 10^{-3}$ (dashed). For the positive slope distribution, the kinematic edge $s_0/m_B^2 = 2.88$ clearly emerges as $\Gamma_B/m_B \to 0$, while the negative slope is always precisely zero at the kinematic edge.}
\label{fig:SIF}
\end{figure}

A few remarks about Eqs. (\ref{eqn:IFDDRT}) and (\ref{eqn:IFNF}) are in order. First, obviously the linear $s$ dependence of the arctangent prefactors in Eqs. (\ref{eqn:IFNF}) changes the overall shape of the chiral differential decay rates from a rectangle into a triangle of either positive or negative slope. Second, just as for the intermediate scalar case, the arctangent functions produce a kinematic edge at $s= s_0$ which is smeared by non-zero $\Gamma_B$. Note, however, that the negatively sloped differential decay rate is precisely zero at the kinematic edge. Finally, from Eqs. (\ref{eqn:IFNF}) and (\ref{eqn:IFDDRT}), we see that in $d\Gamma/ds|_{\rm f}$ the linear $s$ dependent piece of the arctangent prefactor is proportional to
\begin{equation}
	\big[(g_L^X\tilde{g}_L^Y)^2 + (g_R^X\tilde{g}_R^Y)^2\big]-\big[(g_L^X\tilde{g}_R^Y)^2 + (g_R^X\tilde{g}_L^Y)^2\big]~.
\end{equation}
This $s$ dependence disappears if either the $X$ or $Y$ vertex has a vectorial coupling, i.e $g^X_L=g^X_R$ or $g^Y_L=g^Y_R$ respectively. In this case we obtain a square envelope for the differential decay rate. In other words, if at least one coupling is vectorial then there are no spin correlations between $X$ and $Y$, provided one averages over their initial and final spins. This means that for such vectorial coupling, one cannot distinguish a single intermediate fermion from an intermediate scalar using the shape of the differential decay rate. This is a well-known result (see e.g. \cite{Wang:2008sw}). Plots of the chiral differential decay rates are shown in Fig. \ref{fig:SIF}.

\subsection{Intermediate Fermion with Flavor Oscillation}
\begin{figure}[t]
\includegraphics{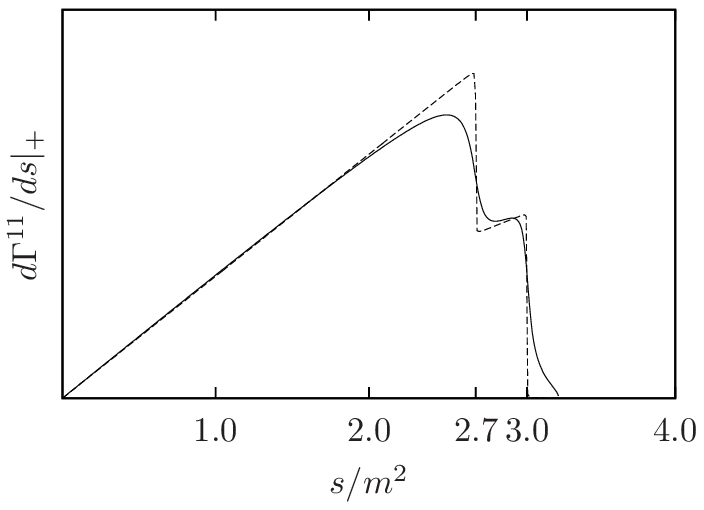}
\includegraphics{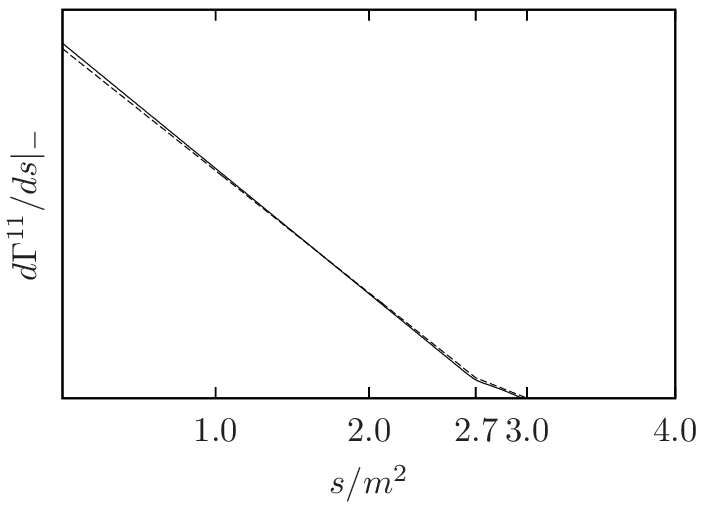}
\includegraphics{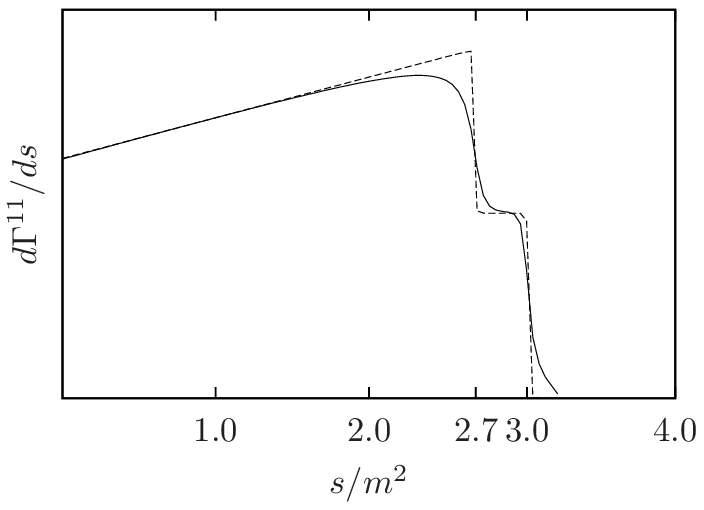}
\includegraphics{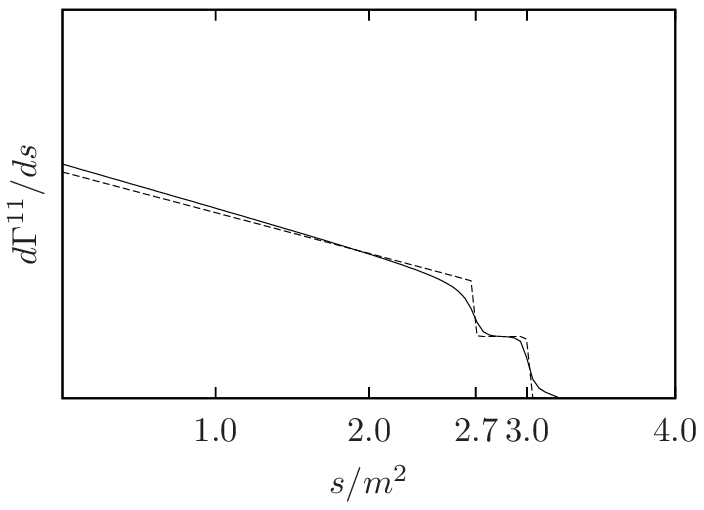}
\caption{Internal fermion chiral differential decay rates $d\Gamma^{11}/ds|_{+}$ (upper left),  $d\Gamma^{11}/ds|_{-}$ (upper right), and summed differential decay rates $d\Gamma^{11}/ds$ (lower left and right). The parameter choices are $\theta = \pi/4$, $m_A/m = 2$, $m_C/m = 0.2$, $z = y = 0.1$, and $x = 4$ (solid) or $x = 4\times10^2$ (dashed), which correspond to $\epsilon = 5\times10^{-2}$ and $5\times10^{-4}$ respectively. For the summed rates we choose couplings such that $(g^X_L\tilde{g}^Y_L)^2 + (g^X_R\tilde{g}^Y_R)^2 = 1/2$, $(g^X_L\tilde{g}^Y_R)^2 + (g^X_R\tilde{g}^Y_L)^2 = 3/2$ (lower left) and vice-versa (lower right). The kinematic edges in $d\Gamma/ds|_{+}$ at $s^1_0/m^2 = 2.7$ and $s_0^2/m^2 = 3.0$ are evident as $x$ becomes large, while in $d\Gamma/ds|_{-}$ we see instead two kinks. The trapezoidal shape and kinematic edges of the summed rates $d\Gamma^{11}/ds$ are manifest.}\label{fig:IFDDR}
\end{figure}

Let us next consider the differential decay rate for the amplitude in Fig. \ref{fig:SFD}b with two-flavor mixing. This time there are two intermediate fermions $B_1$ and $B_2$, with Yukawa-type vertices
\begin{equation}
\mathcal{L}^{\textrm{f}}= \bar{\psi}_B^i\big(g^X_LP_L + g^X_RP_R\big)U^{\alpha i *}\ell^\alpha \phi^\dagger_A + \bar{\psi}_B^i\big(g^Y_LP_L + g^Y_RP_R\big) U^{\alpha i *}\ell^\alpha\phi^\dagger_C~.\label{eqn:LFM}
\end{equation}
Just as for the single intermediate fermion, the differential decay rate is of the form
\begin{equation}
	\label{eqn:IFDDRT2}
	\frac{d\Gamma^{\alpha\beta}}{ds}\bigg|_{\textrm{f}} = \big[(g_L^X\tilde{g}_L^Y)^2 + (g_R^X\tilde{g}_R^Y)^2\big]\frac{d\Gamma^{\alpha\beta}}{ds}\bigg|_- + \big[(g_L^X\tilde{g}_R^Y)^2 + (g_R^X\tilde{g}_L^Y)^2\big]\frac{d\Gamma^{\alpha\beta}}{ds}\bigg|_+~.
\end{equation}
Again the subscript `f' denotes an internal fermion. The chiral differential decay rates, $d\Gamma/ds|_{\pm}$, now have both square and interference terms, so we write
\begin{equation}
	\label{eqn:IFCRD}
	\frac{d\Gamma^{\alpha\beta}}{ds}\bigg|_{\pm} = \frac{d\Gamma^{\alpha\beta}_{1}}{ds}\bigg|_{\pm} + \frac{d\Gamma^{\alpha\beta}_{2}}{ds}\bigg|_{\pm} + \frac{d\Gamma^{\alpha\beta}_{12}}{ds}\bigg|_{\pm}~.
\end{equation}
The square terms are similar to Eqs. (\ref{eqn:IFNF}).  To leading order in $\epsilon$, they are
\begin{align}
	\frac{d\Gamma^{\alpha\beta}_{j}}{ds}\bigg|_{+} & = \frac{|U^{\alpha j}|^2|U^{\beta j}|^2}{(2\pi)^3}\bigg[\frac{s}{m_A^3}\bigg]\bigg(\frac{1 + (-)^jz}{1 + (-)^jy}\bigg)\tan^{-1}\bigg[\frac{\eta}{\epsilon_j}\bigg]\Bigg|_{\eta=\eta^j_-}^{\eta = \eta^j_+}~,\notag\\
	\frac{d\Gamma^{\alpha\beta}_{j}}{ds}\bigg|_{-} & =  \frac{|U^{\alpha j}|^2|U^{\beta j}|^2}{(2\pi)^3}\bigg[\frac{s_0^j-s}{m_A^3}\bigg]\bigg(\frac{1 + (-)^jz}{1 + (-)^jy}\bigg)\tan^{-1}\bigg[\frac{\eta}{\epsilon_j}\bigg]\Bigg|_{\eta=\eta^j_-}^{\eta = \eta^j_+}~.\label{eqn:IFST}
\end{align}
The interference terms are similar to Eq. (\ref{eqn:DDRIS})
\begin{equation}
	\frac{d\Gamma^{\alpha\beta}_{12}}{ds}\bigg|_{\pm}  = \frac{m^2}{(2\pi)^3m_A^3}\bigg[\frac{U^{\alpha 1}U^{\beta 1}U^{\alpha 2}U^{\beta 2}}{x^2 + 1}\bigg]\sum_{j=1,2}\Bigg\{\mathcal{A}_{\pm}^j\tan^{-1}\bigg[\frac{\eta}{\epsilon_j}\bigg] +\mathcal{B}_\pm^j\log\Big[\eta^2 +\epsilon_j^2\Big]\Bigg\}\Bigg|_{\eta=\eta^j_-}^{\eta = \eta^j_+}~,\label{eqn:IFIT}
\end{equation}
except that the coefficients (to leading order in $\epsilon$) are
\begin{align}
	\mathcal{A}^{j}_{+} & = \frac{s}{m^2}~,\notag\\
	\mathcal{B}^{j}_{+} & = -(-)^j\frac{x}{2}\frac{s}{m^2}~,\notag\\
	\mathcal{A}^{j}_{-} & = \mathcal{A}^{j}_{\textrm{sc}} - \frac{s}{m^2}~,\notag\\
	\mathcal{B}^{j}_{-} & = \mathcal{B}^{j}_{\textrm{sc}} + (-)^j \frac{x}{2} \frac{s}{m^2}~. \label{eqn:IFIC}
\end{align}
We have again dropped the $\mathcal{O}(z)$ terms, as discussed in Sec. \ref{sec:ISFM}. Notice that the square terms $d\Gamma_{j}/ds|_{-}$ are precisely zero at their kinematic edges and thereafter. As a result, rather than a double edge, the $d\Gamma/ds|_{-}$ rate has two kinks, which are smeared as $\epsilon$ increases. Plots of the chiral differential decay rates $d\Gamma/ds|_{\pm}$ and summed differential decay rates (\ref{eqn:IFDDRT}) are presented in Fig. \ref{fig:IFDDR}.

Our analysis in Sec.~\ref{sec:ISFM} concerning the role of
the logarithms, arctangents and $1/(x^2 + 1)$ factors in the
interference terms for scalars applies equally to our results for the intermediate
fermions. As mentioned above, the main difference between the
intermediate scalar and intermediate fermion results
are the linear $s$ prefactors, along with the fact that there are
twice as many terms corresponding to the two chiral coupling
cases. Finally, note that in the case a coupling is vectorial, both the
square and interference terms reduce to the intermediate scalar case:
The linear $s$ dependence in the prefactors manifestly cancels and the
fermion and scalar couplings coincide, just as for a single mass
eigenstate. 

This cancellation does not, however, persist in the log interference
terms at higher $\epsilon$ order (see Appendix \ref{sec:AIF} for
details). This non-cancellation may be an artifact of the Breit-Wigner approximation \cite{Badalyan:1982rc,Chung:1995pw} and is certainly negligible here. Yet, it is worth pointing out that such a difference could open possibilities for intermediate particle spin determination in the vectorial coupling case.  A possible physical origin of the difference between the bosonic and the fermionic cases may be off-shellness effects in the interference term: When the
intermediate particles are far off-shell, e.g. in the $m_{1,2}\gg m_A> m_C$ regime,
the differential decay rate is very different depending on the spin of
the intermediate particle. Since off-shellness is parametrized by $\epsilon$, to compute such off-shell spin effects would require analysis of higher order $\epsilon$ terms. This is beyond the scope of this paper, and thus we do not exploit this possible avenue for spin determination any further here.

\section{Observables}
\label{sec:OSD}
Having presented the differential decay rates for the intermediate fermion and scalar cases with flavor oscillation in Sec. \ref{sec:FM}, we now examine what physical observables can in principle be measured from these distributions. We emphasize that we do not study the feasibility of such measurements. A detailed analysis of the physical information contained in the differential decay rates for the limit $x \to \infty$, in which oscillation and widths are negligible, has been conducted in Ref. \cite{Galon:2011wh}. 

For the purposes of this discussion, in this section we will focus on the canonical
example of cascade decay with an intermediate slepton (that is, a scalar)
\begin{equation}
	\tilde{\chi}_1^0\to\ell\tilde{\ell}\to\ell\ell\tilde{\chi}_2^0~.
\end{equation}
In line with our discussion so far, we shall assume that there are only two flavors of slepton which mix significantly. We shall also assume they couple to the neutralino $\chi_{1}$ ($\chi_2$) and the electron $e$ (muon $ \mu$) via CP preserving interactions of the form in Eq. (\ref{eqn:LSF}). We adopt the notation
\begin{equation}
	\psi_{A,C} \equiv \chi_{1,2}~,\qquad \ell^{1,2} \equiv e,\mu~.
\end{equation}
In contrast to the notation defined in Eq. (\ref{eqn:DU}), for the sake of clarity in this section we will henceforth assign the flavor indices of the mixing matrix to be $\alpha = e,\mu$. 

The SUSY spectrum depends on the mechanism of SUSY breaking. If the breaking is mediated in a flavor-universal manner then the mass
splitting is small. For example, gauge mediation in SUSY breaking
theories naturally gives mass splittings and decay widths for sleptons
of the order GeV. Since existing bounds on the slepton masses imply
$m_{\tilde{\ell}} > 10^2$ GeV, then for this decay we consider
oscillation parameters
\begin{equation}
	\label{eqn:MPE}
	m \sim 10^2~\mbox{GeV}~,\qquad y \simeq 0~,\qquad z \sim 10^{-2}~,\qquad x \sim 1~.
\end{equation}
These parameters satisfy the small width condition $\epsilon_{1,2}
\ll1$, and flavor oscillation is strong.

\subsection{Parameter Counting: Kinematic Edges}
As explained in Sec. \ref{sec:MP}, measurement of the four oscillation
parameters $m$, $x$, $y$ and $z$ uniquely determines the slepton
masses and widths. The other physical parameters we wish to measure
are the mixing angle $\theta$, the neutralino masses $m_{\chi_1}$ and
$m_{\chi_2}$, and the couplings $g_{L,R}^X$ and $g_{L,R}^Y$, so
we have eleven physical parameters of interest in total. However, from
Eq. (\ref{eqn:ISDDR}) only the combination
\begin{equation}
	\tilde{g} \equiv \big[(g_L^X)^2 + (g_R^X)^2\big]\big[(\tilde{g}_L^Y)^2 + (\tilde{g}_R^Y)^2\big]~,
\end{equation}
appears in the differential decay rate, so only $\tilde{g}$ can be measured. Thus we have an eight dimensional parameter space.

In the case of two-flavor mixing and with CP conservation, the
differential decay rates are symmetric in flavor indices, as can be
seen from Eqs. (\ref{eqn:STIS}) and (\ref{eqn:DDRIS}). That is,
$d\Gamma^{\alpha\beta}/ds = d\Gamma^{\beta\alpha}/ds$. As a result,
there are three independent differential decay rate distributions
which can be measured, namely
\begin{equation}
	\label{eqn:IDDRD}
	\frac{d\Gamma^{ee}}{ds}~,\qquad \frac{d\Gamma^{e\mu}}{ds} = \frac{d\Gamma^{\mu e}}{ds}~,\qquad\frac{d\Gamma^{\mu\mu}}{ds}~.
\end{equation}
Note that in the most general case $U$ can be complex, and the
interference prefactors then generally contain $U^{\alpha i}U^{\beta i
*}U^{\alpha j*}U^{\beta j}$. As a result $d\Gamma^{\alpha\beta}/ds
\not= d\Gamma^{\beta\alpha}/ds$, and then each final state provides an
independent differential decay rate.

The positions of the kinematic edges (\ref{eqn:KEFM}) are obviously lepton flavor independent, and therefore are the same for these three distributions. The edges therefore yield two independent constraints on the four-dimensional parameter subspace $\{m_{\chi_1}, m_{\chi_2}, m,z\}$, constraining it to a two-dimensional surface. The full parameter space is constrained by the edges to a six-dimensional surface. We now must seek other observables to further constrain the parameter space.

\subsection{Direct Measurement of the Widths and Oscillation}
In Sec. \ref{sec:EW} we established that the smearing of the edges is characterized by the edge width $\sigma$, defined in Eq. (\ref{eqn:DEW}). The edge width approximates the full width at half maximum of the derivative of the differential decay rate, $d\Gamma^2/ds^2$, and is therefore a measurable quantity. Assuming that $1 - m_{\chi_1}^2m_{\chi_2}^2/m^4 \sim \mathcal{O}(1)$ and since $y \simeq 0$, $z\sim 10^{-2} \ll1$, then for the intermediate slepton cascade the edge width reduces simply to
\begin{equation}
	\label{eqn:EWE}
	\sigma \simeq 2\epsilon\frac{m^4 - m_{\chi_1}^2m_{\chi_2}^2}{m^2} = \frac{4m^2z}{x}\bigg(1 - \frac{ m_{\chi_1}^2m_{\chi_2}^2}{m^4}\bigg)~,
\end{equation}
for both kinematic edges. Since this is lepton flavor independent, all three differential decay rates feature two kinematic edges with the same edge width, yielding one independent further constraint on the parameter space. 

Significantly, since the magnitudes of $m$ and $z$ are fixed by other
theoretical considerations (as in Eq. (\ref{eqn:MPE})), and provided
$1 - m_{\chi_1}^2m_{\chi_2}^2/m^4 \sim \mathcal{O}(1)$, then
measurement of $\sigma$ provides the magnitude of $x$. As
a consequence, the measurement of the smearing of the edges  --- the
edge width --- measures the degree of flavor oscillation. 

\subsection{Edge Resolution Criterion}
\label{sec:ERC}
\begin{figure}[t]
\includegraphics{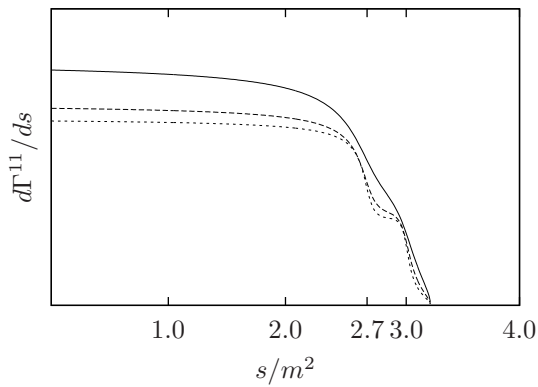}
\caption{Intermediate scalar differential decay rate, with the parameter choices $\theta = \pi/4$, $m_A/m = 2$, $m_C/m = 0.2$, $y=0$, $z = 0.1$, and $x = 1$ (solid), $x = 2$ (dashed) or $x=3$ (dotted). These correspond to $\epsilon = 0.2$, $\epsilon = 0.1$ and $\epsilon = 0.02$ respectively. }
\label{fig:ERC}
\end{figure}

So far in this discussion we have assumed that it is possible to distinguish the two edges, that is to say, it is possible to resolve them. But if the separation of the edges is of similar or smaller size than their widths, $\sigma_{1,2}$, then it is reasonable to assume that one may not resolve the two edges.  This leads to a natural edge resolution criterion. For two kinematic edges to be resolvable, we require the separation of the edges to be greater than their average width, that is
\begin{equation}
	\big|s_0^1 - s_0^2\big| > \frac{\sigma_1 + \sigma_2}{2}~.
\end{equation}
For $z \ll 1$ and $1 - m_{\chi_1}^2m_{\chi_2}^2/m^4 \sim \mathcal{O}(1)$, this criterion reduces to just a simple restriction on $x$, 
\begin{equation}
	\label{eqn:ASERC}
	x > 1~.
\end{equation}
This is an interesting result, because the resolvability depends on both the separation and the widths of the edges, so na\"\i vely we would expect the resolution criterion to depend on both $x$ and $z$. That said, this result implies that oscillation and edge resolvability are inversely correlated, which aligns with our expectation that the edges should be resolvable when interference is negligible. For the intermediate slepton cascade we have $x \sim 1$, so the edges may not always be resolvable, but they are if we restrict our attention to $x > 1$, which is often the regime of interest. A graphical demonstration of the sensitivity of this criterion is provided in Fig. \ref{fig:ERC}. For $x = 2$ the edges are already visibly resolvable. Note that the parameters have been chosen in this figure to match our prior discussion, while still satisfying $z,\epsilon \ll1$ and $1 - m_{\chi_1}^2m_{\chi_2}^2/m^4 \sim \mathcal{O}(1)$, and do not arise from the slepton cascade.

\subsection{Edge Degeneracy and the Geometric Mean}
Before continuing, we wish to point out that kinematic edges can be irresolvable even in the absence of strong oscillation.  That is, there are special points in parameter space at which two kinematic edges (\ref{eqn:KEFM}) are degenerate, i.e. $s_0^{1} = s_0^2$. This occurs not only for the obvious case that $m_1 = m_2$, but also if
\begin{equation}
	\label{eqn:EDC}
	m_1m_2 = m_Am_C~,
\end{equation}
that is, if the geometric means of $m_{1,2}$ and $m_{A,C}$ are the same. In terms of the parameter $z$, this condition corresponds to 
\begin{equation}
	z = \sqrt{1 - \frac{m_Am_C}{m^2}}~.
\end{equation}
If we require $z \ll 1$, then it follows that $1 - m_{\chi_1}^2m_{\chi_2}^2/m^4 \ll 1$, too. So Eq. (\ref{eqn:EDC}) is satisfied in a different regime of parameter space to that considered in Sec. \ref{sec:ERC}. 

The physical origin of this degeneracy can be understood by observing that if $X$ and $Y$ are massless, then the invariant mass $s = 2(|\vec{p}_X||\vec{p}_Y| - \vec{p}_X\cdot\vec{p}_Y) \le 4|\vec{p}_X||\vec{p}_Y|$.  One derives the kinematic edge, $s^j_0$, from kinematics by maximizing $s$ subject to $B_j$ being on-shell. It is clear that a given value of $s$ can be obtained from many configurations of $\vec{p}_X$ and $\vec{p}_Y$. For instance, $s$ has the symmetry $|\vec{p}_X| \to \lambda|\vec{p}_Y|$, $|\vec{p}_Y| \to |\vec{p}_X|/\lambda$, $\forall\lambda\in\mathbb{R}$. 
It is therefore not surprising that there exist different on-shell $B$ masses with correspondingly different $p_{X,Y}$ configurations, but with $s_0$ the same. In particular, it can be shown that $m_1 \to m_Am_C/m_1= m_2$ is equivalent to $\lambda = m_A/m_C$. So if $m_1m_2 = m_Am_C$, then the two kinematic edges are degenerate.

When Eq. (\ref{eqn:EDC}) is satisfied, the kinematic edges collide, so
that one may not determine the number of intermediate mass eigenstates
from the differential decay rate $d\Gamma/ds$. There may, of course,
still be oscillation between the two $B$'s in this case or none at all.  The moral of this
section and Sec. \ref{sec:ERC} is that the observation of a certain
number of kinematic edges provides only a lower bound on the number of
intermediate degrees of freedom.

\subsection{Step Height Ratios}
If the edges are resolvable, the ratio of the heights of the step functions that form the kinematic edges is another observable. For $y \simeq 0$ and $z \ll1$, using the results of Sec. \ref{sec:ISFM} one may show that this step height ratio is to leading order in $z$
\begin{equation}
	\label{eqn:RAB}
	R^{\alpha\beta} = \frac{|U^{\alpha 1}|^2|U^{\beta 1}|^2 + U^{\alpha 1}U^{\beta 1}U^{\alpha 2}U^{\beta 2}[1 + (x/\pi)\log\epsilon]/(x^2 +1)}{|U^{\alpha 2}|^2|U^{\beta 2}|^2 + U^{\alpha 1}U^{\beta 1}U^{\alpha 2}U^{\beta 2}[1 - (x/\pi)\log\epsilon]/(x^2 +1)}~.
\end{equation}
The step height ratio is lepton flavor dependent, yielding three constraints on the three dimensional $\{z,x,\theta\}$ parameter space.  Consequently, the step height ratios probe $z$, $x$ and $\theta$. Reversing the parameter counting above, measurement of $\sigma$ now constrains $m$ uniquely, while the two kinematic edges are left to constrain $m_{\chi_1}$ and $m_{\chi_2}$. With $y \simeq 0$, only the coupling combination $\tilde{g}$ is left unconstrained by these observables.

\subsection{$s=0$ Intercepts}
We now consider the value of the decay rate at $s=0$, which we call the $s=0$
intercept. These intercepts are another three physical
observables. Since they depend on overall normalization, they permit
measurement of $\tilde{g}$, and moreover, they are measurable even if
the edges cannot be resolved. Explicitly, for the three independent
differential decay rates, these intercepts are to leading order
\begin{equation}
	\label{eqn:ASI}
	I^{\alpha\beta} = \tilde{g}\frac{(m_{\chi_1}^2 - m^2)(m^2 - m_{\chi_2}^2)}{8\pi^2m_A^3m^2}\bigg[|U^{\alpha 1}|^2|U^{\beta 1}|^2 + |U^{\alpha 2}|^2|U^{\beta 2}|^2+ \frac{2}{x^2+1}U^{\alpha 1}U^{\beta 1}U^{\alpha 2}U^{\beta 2}\bigg]~.
\end{equation}
Unlike the step height ratios, not only does $I^{e\mu} = I^{\mu e}$, but also $I^{ee} = I^{\mu\mu}$, so the intercepts actually provide just two independent observables. The kinematic edges, edge width, step height ratios and $s=0$ intercepts thus provide eight physical observables for the original eight-dimensional parameter space. Alternatively, if we set $y\simeq0$, then the seven-dimensional parameter space is over-constrained. Note also that the single independent \emph{ratio} of these intercepts is normalization independent, and therefore constrains just the $\{x, \theta\}$ parameter subspace. 

\subsection{Intermediate Fermion Observables}
Let us now consider the physical observables for the case of  two-flavor mixing with intermediate fermions. It is convenient to define
\begin{equation}
	\tilde{g}_+ \equiv (g_L^X\tilde{g}_L^Y)^2 + (g_R^X\tilde{g}_R^Y)^2~,\qquad \tilde{g}_- \equiv (g_L^X\tilde{g}_R^Y)^2 + (g_R^X\tilde{g}_L^Y)^2~,
\end{equation} 
which are the measurable coupling combinations analogous to $\tilde{g}$ above. The parameter space for fermions is therefore enlarged by one dimension compared to the scalar case. We assume the same oscillation parameter structure as in Eqs. (\ref{eqn:MPE}).

Just as for the intermediate scalar case above, there are still three independent differential decay distributions (\ref{eqn:IDDRD}); the kinematic edges and edge width still provide respectively two and one physical observables that constrain the parameter space; and the edge resolution criterion (\ref{eqn:ASERC}) is the same. Apart from the dimensionalities of their parameter spaces, the only other difference between the scalar and fermionic cases is that for the latter, rather than step height ratios one must instead measure the ratio of trapezoid apex heights. That is, one must extrapolate the smeared edges into sharp corners --- the solid to the dotted lines in Fig. \ref{fig:IFDDR} --- and measure the ratio of the trapezoid apex heights so formed. This is obviously possible, but will presumably introduce more error into the experimental results. At leading order in $\epsilon$ and $z$ these apex height ratios are identical to the step height ratios in Eq. (\ref{eqn:RAB}), and the leading order contribution is due to $d\Gamma/ds|_+$ only.

The $s=0$ intercepts are similarly the same as in Eq. (\ref{eqn:ASI}) but with $\tilde{g}$ replaced by just $\tilde{g}_-$, since $d\Gamma/ds|_+$ is manifestly zero at $s=0$, so the intercepts depend only on $\tilde{g}_-$. One can constrain $\tilde{g}_+$ by measuring the apex heights themselves, rather than their ratio, which depend only on $\tilde{g}_+$ to leading order in $z$.

\section{Conclusion}
\label{sec:C}
In this paper we have presented explicit differential decay rates for
cascade decays of the form (\ref{eqn:CD}) as a function of the $XY$
invariant mass, $s$, including the effects of both finite particle
widths and flavor oscillation. In particular, we considered both scalar and
spin-1/2 intermediate particles with Yukawa-type vertices. Our results
successfully reproduce the usual kinematic edge results in zero width,
zero oscillation limit, and we have shown how to quantify the degree of
kinematic edge smearing for finite widths.

The main results of this paper, however, involve the analysis of the
interplay between non-zero width and flavor oscillation effects, as
characterized by the parameters $x \equiv \Delta m /\bGamma$ and $z
\equiv \Delta m/2m$. Not only does $x$ control the degree of oscillation or
interference between the two $B$ mass eigenstates, it also plays an
important role in the degree of smearing of the two kinematic edges
and their resolvability. In general, the larger $x$, the smaller the
smearing and interference, and the greater the resolvability.

We have also shown how various physical observables of the differential decay rates can be used to constrain the oscillation parameter space, mainly for the special case of a slepton cascade with parameters motivated by gauge mediation SUSY breaking theories. In particular, apart from the kinematic edges, these observables include the edge widths, step height ratios and the $s=0$ intercepts.  Building on the special case considered in Sec. \ref{sec:OSD}, a subject of future work may be to apply our explicit results to more realistic, more complicated or more general scenarios.

Another avenue of study may be to extend the current two-flavor treatment to the three or more flavors. Furthermore,  one might lift the CP conservation assumption which we have made throughout this paper. Such CP violation will produce CP violating interference terms, which may have interesting physical effects. We plan to address these issues in the future.

\acknowledgments
We thank Josh Berger, David Curtin, Iftah Galon, Maxim Perelstein and Yael Shadmi
for helpful discussions. This work is supported by the U.S. National
Science Foundation through grant PHY-0757868.

\appendix

\section{Phase Space Integral and Non-Zero Width}
\label{sec:AppA}

In the following we compute the differential decay rate $d\Gamma/ds$ for a cascade decay of the general form (\ref{eqn:CD})
\begin{equation}
	\label{eqn:CDA}
	A \to XB \to XYC~.
\end{equation}
Although some of what is reported in this appendix is already well-known, it is our hope that its recapitulation here is a useful reference to the reader.

For the three body decay (\ref{eqn:CDA}) the differential decay rate is \cite{Peskin:1996,Weinberg:1996vol1}
\begin{equation}
\label{eqn:DDR}
d\Gamma =\frac{|\mathcal{M}|^2}{2(2\pi)^5 m_A}\frac{d^3\vec{p}_X}{2E_X}\frac{d^3\vec{p}_Y}{2E_Y}\frac{d^3\vec{p}_C}{2E_C}\delta^{(4)}(p_A-p_X-p_Y-p_C)~,
\end{equation}
where $E_j=(p_j^2+m_j^2)^{1/2}$ and $|\mathcal{M}|^2$ is the amplitude-squared, to be specified later. The differential decay rate is obviously defined in the $A$ rest frame. The amplitude-squared $|\mathcal{M}|^2$ is averaged over all spin indices (if any). As a result, note that $|\mathcal{M}|^2$ has spherical symmetry under arbitrary spatial rotations of the external momenta configuration. We consider only amplitudes of form
\begin{equation}
	\label{eqn:AMSS}
	|\mathcal{M}|^2 = |\mathcal{M}|^2(s,s_1)~,
\end{equation}
where $s = (p_X + p_Y)^2$ and $s_1 = (p_A - p_X)^2$. 

To obtain the differential decay rate, $d\Gamma/ds$, we must integrate out all other variables except for the Lorentz invariant $s = (p_X + p_Y)^2$. To do this, we first integrate out $\vec{p}_Y$ using the momentum conserving delta function. Changing to polar coordinates, we then have
\begin{equation}
	\label{eqn:DDR2}
	\Gamma =\frac{1}{2(2\pi)^5m_A}\int\frac{|\vec{p}_X|^2d|\vec{p}_X|d\Omega_X|\vec{p}_C|^2d|\vec{p}_C|d\Omega_C}{8E_XE_YE_C}|\mathcal{M}|^2\delta(m_A-E_X-E_Y-E_C)~.
\end{equation}
By Eq. (\ref{eqn:AMSS}), the choice for the axes with respect to which $\Omega_X$ and $\Omega_C$ are calculated is arbitrary.  We choose the 3-axis to coincide with the direction of $\vec{p}_X$ so that
\begin{equation}
	\label{eqn:AAD}
	d\Omega_C=d\cos{\theta_{XC}}\ d\phi_C~,
\end{equation}
where $\theta_{XC}$ is the angle between $\vec{p}_C$ and $\vec{p}_X$. Exploiting the simple relation
\begin{equation}
	\label{eqn:AEP}
	E_idE_i=|\vec{p}_i|d|\vec{p}_i|
\end{equation}
and performing the trivial $d\Omega_X$ and $d\phi_C$ integrals, Eq. (\ref{eqn:DDR2}) reduces to
\begin{equation}
	\Gamma =\frac{1}{(2\pi)^3m_A}\int\frac{|\vec{p}_X||\vec{p}_C|dE_XdE_Cd\cos{\theta_{XC}}}{8E_Y}|\mathcal{M}|^2\delta(m_A-E_X-E_Y-E_C)~.
\end{equation}
Note that the $d\cos{\theta_{XC}}$ integral is non-trivial, since $E_Y$ is a function of $\cos{\theta_{XC}}$. That is,
\begin{equation}
	\label{eqn:AEYC}
	E_Y=\sqrt{|\vec{p}_X+\vec{p}_C|^2+m_Y^2}=\sqrt{m_X^2+m_C^2+m_Y^2+2|\vec{p}_X||\vec{p}_C|\cos{\theta_{XC}}}~.
\end{equation}

We are left with three integrations and one delta function. First, to conveniently encode the $[-1,1]$ domain of the $\cos{\theta_{XC}}$ integral, we add a theta function factor, $\Theta(1 - \cos{\theta_{XC}}^2)$ and extend the $\cos\theta_{XC}$ integration limits to the entire real line. We now use the delta function to perform the integration over $d \cos{\theta_{XC}}$. From Eq. (\ref{eqn:AEYC}) observe that
\begin{equation}
	\frac{dE_Y}{d\cos{\theta_{XC}}}=\frac{|\vec{p}_X||\vec{p}_C|}{\sqrt{m_X^2+m_C^2+m_Y^2+2|\vec{p}_X||\vec{p}_C|\cos{\theta_{XC}}}}=\frac{|\vec{p}_X||\vec{p}_C|}{E_Y}~,
\end{equation}
so that we may straightforwardly change variable from $\cos{\theta_{XC}}$ to $E_Y$ and perform the final delta function integral. We end up with
\begin{equation}
\label{eqn:DDR3}
\Gamma_{A\to X,Y,C}=\frac{1}{8(2\pi)^3m_A}\int dE_XdE_C|\mathcal{M}|^2\Theta(1-\cos^2{\theta_{XC}})~.
\end{equation}
Here $\cos\theta_{XC}$ is implicitly a function of $E_{X,C}$ and the masses. 

We can now change variable from $(E_X,E_C)$ to $(s,s_1)$ using
\begin{equation}
\label{eqn:AXCSS}
	E_X=\frac{m^2_A+m_X^2-s_1}{2m_A}, \qquad E_C=\frac{m^2_A+m_C^2-s}{2m_A}\qquad\Rightarrow
	\quad dE_XdE_C=\frac{dsds_1}{4m_A^2}~,
\end{equation}
then from Eq. (\ref{eqn:DDR3})
\begin{equation}
	\frac{d\Gamma}{ds}=\frac{1}{32(2\pi)^3m^3 _A}\int ds_1|\mathcal{M}|^2(s,s_1)\Theta(1-\cos^2{\theta_{XC}})~.
\end{equation}
In order to obtain the differential decay rate we need to perform just the $s_1$ integration. This integral can be performed once an explicit expression for $|\mathcal{M}|^2$ is given.

Before proceeding to consider the differential decay rates arising from different matrix elements, we still need to explicitly write down the limits of the $s_1$ integration due to the $\Theta$ function. In order to do so we must write $\cos{\theta_{XC}}$ as a function of $(s,s_1)$ and then solve $\cos^2{\theta_{XC}}\leq1$ for $s_1$. Define $s_2=(p_X+p_C)^2$. Then
\begin{equation}
	\label{eqn:S1}
	s_2=(p_X+p_C)^2=m_X^2+m_C^2+2E_XE_C-2|\vec{p}_X||\vec{p}_C|\cos{\theta_{XC}}~.
\end{equation}
Applying the identities (\ref{eqn:AXCSS}), plus
\begin{equation}
	|\vec{p}_i|=\sqrt{E^2_i-m_i^2},\qquad s + s_1 + s_2=m_X^2+m_C^2+m_A^2+m_Y^2~,
\end{equation}
Eq. (\ref{eqn:S1}) becomes
\begin{equation}
	\label{eqn:ACXC}
	\cos{\theta_{XC}}=\frac{2m_A^2(s+s_1-m_Y^2-m_A^2)+(m_A^2+m_X^2-s_1)(m_A^2+m_C^2-s)}{\sqrt{(m^2_A+m_C^2-s)^2-4m_A^2m_C^2}\sqrt{(m_A^2+m_X^2-s_1)^2-4m_A^2m_X^2}}~.
\end{equation}

The general solution for $\cos^2\theta_{XC}\leq1$ in terms of $s_1$ is $s_{1_-} \le s_1 \le s_{1_+}$ with limits
\begin{align}
	s_{1_\pm}
	& = \frac{m_A^2 + m_X^2 + m_Y^2 + m_C^2 - s}{2} + \frac{(m_Y^2 -m_X^2)(m_A^2 - m_C^2)}{2s}\notag\\
	& \quad \pm \frac{2}{s}\sqrt{\bigg[\bigg(\frac{m^2_A+m^2_C- s}{2}\bigg)^2 - m_A^2m_C^2\bigg]\bigg[\bigg(\frac{m^2_Y- m^2_X+ s}{2}\bigg)^2 - m_Y^2s\bigg]}~.\label{eqn:S1PM}
\end{align}
We now have our master formula for the differential decay rate
\begin{equation}
\label{eqn:FDDR}
	\frac{d\Gamma}{ds}=\frac{1}{32(2\pi)^3m^3_A}\int^{ s_{1_+}}_{s_{1_-}}ds_1|\mathcal{M}|(s,s_1)^2~.
\end{equation}
If the two leptons $X$ and $Y$ are massless, the integration limits reduce to
\begin{equation}
	\label{eqn:S1PMM}
	s_{1_{\pm}} = \frac{m^2_A+m^2_C- s}{2} \pm \sqrt{\bigg(\frac{m^2_A+m^2_C- s}{2}\bigg)^2 - m_A^2m_C^2}~.
\end{equation}
Defining a new variable $\eta \equiv 1 - s_1/m_B^2$, we immediately obtain the integration limits $\eta_{\pm}$ and the natural definition of $\xi$ as written in Eq. (\ref{eqn:DEX}) of the main text.

We now consider explicit examples and carry out the integral in (\ref{eqn:FDDR}) for the three cases considered in the main text.

\subsection{$\phi^3$ interaction}

First consider the case in which all the particles involved are scalars with $\phi^3$ vertices of the form
\beq
g_X\phi_A\phi_B\phi_X+g_Y\phi_C\phi_B\phi_Y~.
\eeq
To take into account the decay width $\Gamma_B$ we use the Breit-Wigner approximation described in the main text (\ref{eqn:BWS}). The matrix element is
\beq
|\mathcal{M}|_{\phi^3}^2=\frac{g_X^2g_Y^2}{((p-p_X)^2-m_B^2)^2+m_B^2\Gamma_B^2}=\frac{g_X^2g_Y^2}{(s_1-m_B^2)^2+m_B^2\Gamma_B^2}~,
\eeq
so the master formula (\ref{eqn:FDDR}) then becomes
\beq
\frac{d\Gamma}{ds}=\frac{g_X^2g_Y^2}{32(2\pi)^3m^3_A}\int^{ s_{1_+}}_{s_{1_-}}ds_1\frac{1}{(s_1-m_B^2)+m_B^2\Gamma^2_B}~.
\eeq
Using
\beq
\int ds_1\frac{1}{(s_1-m_B^2)+m_B^2\Gamma^2_B}=\frac{1}{m_B\Gamma_B}\ArcTan\bigg({\frac{s_1-m_B^2}{m_B\Gamma_B}}\bigg)~,
\eeq
we get the final result reported in the main text
\beq
\frac{d\Gamma}{ds}=\frac{g_X^2g_Y^2}{32(2\pi)^3m^3_Am_B\Gamma_B}\bigg[\ArcTan\bigg(\frac{m_B}{\Gamma_B}\eta_+\bigg)-\ArcTan\bigg(\frac{m_B}{\Gamma_B}\eta_-\bigg)\bigg]~,
\eeq
where $\eta_\mp = 1 - s_{1_{\pm}}/m_B^2$ as defined in the main text (cf. Eq. (\ref{eqn:DEX}))~.

\subsection{Intermediate Scalar}
\label{sec:AIS}
We now move on to consider  the decay in Fig. \ref{fig:SFD}a. The corresponding Yukawa couplings are (Eq. (\ref{eqn:LSF}))
\begin{equation}
	\mathcal{L}^{\textrm{s}}_{\textrm{yuk}} = \bar{\psi}_A\big(g^X_LP_L + g^X_RP_R\big)\ell_X^\alpha U^{\alpha i *}\phi^i_B + \bar{\psi}_C\big(g^Y_LP_L + g^Y_RP_R\big)\ell_Y^\alpha U^{\alpha i *}\phi^i_B~,
\end{equation}
which include flavor mixing. We denote the amplitude for the $j$th $B$ mass eigenstate as $\mathcal{M}^{{\rm sc}}_j$, and
\begin{equation}
\label{eqn:AISM}
i\mathcal{M}^{{\rm sc}}_j = \bar{u}^\alpha(p_X)\big[g_L^XP_R + g_R^XP_L\big]u(p_A)\bar{u}(p_C)\big[g_L^YP_L + g_R^YP_R\big] v^\beta(p_Y)\frac{iU^{\alpha j}U^{\beta j*}}{(p_B^2 - m_j^2) + im_j\Gamma_j}~.
\end{equation} 

The total squared amplitude final involves both square and interference terms
\begin{equation}
	\label{eqn:TA}
	|\mathcal{M}_{{\rm Tot}}^{{\rm sc}}|^2 = |\mathcal{M}_1^{{\rm sc}}|^2+|\mathcal{M}_2^{{\rm sc}}|^2+ 2\mbox{Re}\Big[\mathcal{M}_1^{{\rm sc}}(\mathcal{M}_2^{{\rm sc}})^*\Big]~.
\end{equation}
We focus first on the interference term. After computing the traces, we have 
\begin{align}
	2\mbox{Re}\Big[\mathcal{M}_1^{{\rm sc}}(\mathcal{M}_2^{{\rm sc}})^*\Big]&=2\Big[(g^X_L)^2+(g^X_R)^2\Big]\Big[(g^Y_L)^2+(g^Y_R)^2\Big]\mbox{Re}[U_{\alpha i}U_{\beta j}U^*_{\alpha j}U_{\beta i}^*]\notag\\
	&\quad\times\frac{\Big[(s_1-m_1^2)(s_1-m_2^2)+m_1m_2\Gamma_1\Gamma_2\Big]}{\Big[(s_1-m_1^2)^2+m_1^2\Gamma_1^2\Big]\Big[(s_1-m_2^2)^2+m_2^2\Gamma_2^2\Big]}\mathcal {T}^{{\rm sc}}~,\label{eqn:AIT}\\
\end{align}
and
\begin{equation}
	\mathcal{T}^{{\rm sc}} = 64(p_A\cdot p_X) (p_C\cdot p_Y) = 16(m_A^2-s_1)(s_1-m_C^2)~.
\end{equation}
Note that $s_1=p_B^2$. The square terms $|\mathcal{M}^{{\rm sc}}_j|^2$ follow from Eq. (\ref{eqn:AISM})
\begin{equation}
	\label{eqn:AST}
	|\mathcal{M}_j^{{\rm sc}}|^2=16\Big[(g^X_L)^2+(g^X_R)^2\Big]\Big[(g^Y_L)^2+(g^Y_R)^2\Big]|U^{\alpha j}|^2|U^{\beta j}|^2\frac{(m_A^2-s_1)(s_1-m_C^2)}{(s_1-m_j^2)^2+m_j^2\Gamma_j^2}~.
\end{equation}
The differential decay rate can be obtained plugging Eqs. (\ref{eqn:AIT}) and (\ref{eqn:AST}) in the master formula (\ref{eqn:FDDR}). The final result is the sum of three integrations, involving the square terms and the interference one
\begin{equation}
	\frac{d\Gamma^{\alpha\beta}}{ds}\bigg|_{{\rm sc}}=\frac{\Big[(g^X_L)^2+(g^X_R)^2\Big]\Big[(g^Y_L)^2+(g^Y_R)^2\Big]}{(2\pi)^3m^3_A}\bigg\{\sum_{j=1,2}\mathcal{I}_j+\mathcal{I}_{12}\bigg\}~.
\end{equation}
Here, 
\begin{align}
\mathcal{I}_j&=|U^{\alpha j}|^2|U^{\beta j}|^2\int^{s_{1_+}}_{s_{1_-}}ds_1\frac{(m_A^2-s_1)(s_1-m_C^2)}{(s_1-m_j^2)^2+m_j^2\Gamma_j^2}\notag\\
		    &=m^2_j|U^{\alpha j}|^2|U^{\beta j}|^2\bigg[\frac{1+(-)^jz}{1+(-)^jy}\bigg]\Bigg\{\bigg(\frac{s_0^j+\Gamma_j^2}{m^2_j}\bigg)\tan^{-1}\bigg[\frac{m_j}{\Gamma_j}\eta\bigg]\Bigg|^{\eta=\eta^j_+}_{\eta=\eta^j_-}\notag\\
		    &\qquad\qquad+\frac{\Gamma_j}{m_j}\bigg[\bigg(\frac{m_A^2+m_C^2}{2m^2_j}-1\bigg)\log\bigg[\eta^2+\bigg(\frac{\Gamma_j}{m_j}\bigg)^2\bigg]-\eta\bigg]\Bigg|^{\eta=\eta^j_+}_{\eta=\eta^j_-}\Bigg\}\notag\\
		    & \simeq |U^{\alpha j}|^2|U^{\beta j}|^2 s_0^j \bigg[\frac{1+(-)^jz}{1+(-)^jy}\bigg] \tan^{-1}\bigg[\frac{m_j}{\Gamma_j}\eta\bigg]\Bigg|^{\eta=\eta^j_+}_{\eta=\eta^j_-}~,\label{eqn:AIJ}
\end{align}
where $m$, $z$, $y$, and $x$ are the oscillation parameters defined in Eqs. (\ref{eqn:DM}) and (\ref{eqn:DMXYZ}), $\eta^j_{\mp} \equiv 1 - s_{1_{\pm}}/m_j^2$,
\begin{equation}
	s_0^j=\frac{(m_A^2-m_j^2)(m_j^2-m_C^2)}{m^2_j}~,
\end{equation}
and the approximation in Eq. (\ref{eqn:AIJ}) drops terms whose prefactors are suppressed by $\Gamma_j/m_j$.

The interference term is more involved. We report results to leading order in $\epsilon$ and we directly write the coefficients in terms of the oscillation parameters as in the main text. We have
\begin{align}
\mathcal{I}_{12}&=2\mbox{Re}[U_{\alpha 1}U_{\beta 1}^*U^*_{\alpha 2}U_{\beta 2}]\int^{s_{1_+}}_{s_{1_-}}ds_1\frac{\Big[(s_1-m_1^2)(s_1-m_2^2)+m_1m_2\Gamma_1\Gamma_2\Big]
					\mathcal{T}^{{\rm sc}}}{\Big[(s_1-m_1^2)^2+m_1^2\Gamma_1^2\Big]\Big[(s_1-m_2^2)^2+m_2^2\Gamma_2^2\Big]}\notag\\
				&=\frac{\mbox{Re}[U_{\alpha 1}U_{\beta 1}^*U^*_{\alpha 2}U_{\beta 2}]m^2}{x^2+(1+yz)^2}\sum_{j=1,2}\Bigg\{\mathcal{A}^j_{{\rm sc}}\tan^{-1}\bigg[\frac{m_j}{\Gamma_j}\eta\bigg]\Bigg|^{\eta=\eta^j_+}_{\eta=\eta^j_-}\notag\\
				& \quad\quad\quad -\mathcal{B}^j_{{\rm sc}}\log\bigg[\eta^2+\bigg(\frac{\Gamma_j}{m_j}\bigg)^2\bigg]\Bigg|^{\eta=\eta^j_+}_{\eta=\eta^j_-}\Bigg\}~,
\end{align}
where (at leading order in $\Gamma/m=2z/x$) 
\begin{align}
\mathcal{A}^j_{{\rm sc}}&=\frac{m_A^2m^2(1-yz)(1-z^2)-m^2_Am^2_C(1+yz)+m_C^2m^2(1-yz)(1-z^2)}{m^4}\notag\\
		&-\frac{m^4(1+(-)^jz)^3[1+(-)^jz(-3+3y+yz)]}{m^4}~,\notag\\
\mathcal{B}^j_{{\rm sc}}&=\pm \frac{x}{2}\frac{s_0^jm^2_j}{m^4}~.\label{eqn:ASAB}
\end{align}
These coefficients and $\mathcal{I}_{12}$ prefactors reduce to those in Eqs. (\ref{eqn:ITIS}) at leading order in $z$.

\subsection{Intermediate Fermion}
\label{sec:AIF}
Finally, suppose the intermediate particles are fermions. The Yukawa couplings are as in Eq. (\ref{eqn:LF})
\begin{equation}
\mathcal{L}^{\textrm{f}}_{\textrm{yuk}} = \bar{\psi}_B^i\big(g^X_LP_L + g^X_RP_R\big)U^{\alpha i *}\ell_X^\alpha \phi^\dagger_A + \bar{\psi}_B^i\big(g^Y_LP_L + g^Y_RP_R\big) U^{\alpha i *}\ell_Y^\alpha\phi^\dagger_C~.
\end{equation}
We denote the amplitude for the $j$th $B$ mass eigenstate as $\mathcal{M}^{{\rm f}}_j$, and
\begin{equation}
i\mathcal{M}^{{\rm f}}_j = iU^{\alpha j}U^{\beta j*}\bar{u}^\alpha(p_X)\big[g_L^XP_R + g_R^XP_L\big]\frac{\slashed{p}_B + m_j - i\Gamma_j/2}{(p_B^2 - m_j^2) + im_j\Gamma_j}\big[g_L^YP_L + g_R^YP_R\big] v^\beta(p_Y)~.
\end{equation}	

Now, since $A$ and $C$ are scalars, the total helicity of the $|XY\rangle$ final state must be zero. Since $X$ and $Y$ are massless, spin-1/2 particles, if they have the same (opposite) chirality then they must have opposite (same) sign components of momentum in the angular momentum direction. These are the only options. As a result, the terms of the differential decay rate may depend only on a linear combination of $(g^X_Lg^Y_L)^2$, $(g^X_Rg^Y_R)^2$, $(g^X_Rg^Y_L)^2$ and $(g^X_Lg^Y_R)^2$, and the terms for the latter (former) two factors must have the same $s$ dependence. Hence we may write the differential decay rate in the form (cf. Eqs. (\ref{eqn:IFDDRT}) and (\ref{eqn:IFDDRT2}))
\begin{equation}
	\frac{d\Gamma^{\alpha\beta}}{ds}\bigg|_{\textrm{f}} = \big[(g_L^X\tilde{g}_L^Y)^2 + (g_R^X\tilde{g}_R^Y)^2\big]\frac{d\Gamma^{\alpha\beta}}{ds}\bigg|_- + \big[(g_L^X\tilde{g}_R^Y)^2 + (g_R^X\tilde{g}_L^Y)^2\big]\frac{d\Gamma^{\alpha\beta}}{ds}\bigg|_+~.
\end{equation}
Here $d\Gamma/ds|_{\pm}$ are called the chiral decay rates for reasons explained in the main text, and each one of them has both square and interference contributions:
\begin{equation}
\frac{d\Gamma^{\alpha\beta}}{ds}\bigg|_{\pm}=\frac{d\Gamma^{\alpha\beta}_1}{ds}\bigg|_{\pm}+\frac{d\Gamma^{\alpha\beta}_2}{ds}\bigg|_{\pm}+\frac{d\Gamma^{\alpha\beta}_{12}}{ds}\bigg|_{\pm}~.
\end{equation}

For the sake of brevity, we introduce the following notation
\begin{equation}
	g_+=(g_L^Xg^Y_R)^2+(g_R^Xg^Y_L)^2,\qquad g_-=(g_L^Xg^Y_L)^2+(g_R^Xg^Y_R)^2~,
\end{equation}
and simply report here all the integrands in terms of the variables of integration $(s,s_1)$. 
The interference term is
\begin{align}
\mathcal{M}^{{\rm f}}_{12}\big|_+= & \frac{32g_+\mbox{Re}[U_{\alpha 1}U_{\beta 1}^*U^*_{\alpha 2}U_{\beta 2}]}{\big[(s_1-m_1^2)^2+m^2_1\Gamma_1^2\big]\big[(s_1-m_2^2)^2+m_2^2\Gamma_2^2\big]}\notag\\
			&\quad\quad\times \bigg\{m_1m_2s\Big[(s_1-m_1^2)(s_1-m_2^2)+m_1m_2\Gamma_1\Gamma_2\Big]\notag\\
			&\quad\quad\quad+\frac{1}{2}\Big[(s_1-m_2^2)m_1\Gamma_1-(s_1-m_1^2)m_2\Gamma_2\Big](m_1\Gamma_2-m_2\Gamma_1) s\bigg\}~,\label{eqn:IFINT}\\
\mathcal{M}^{{\rm f}}_{12}\big|_-= & \frac{32 g_- \mbox{Re}[U_{\alpha 1}U_{\beta 1}^*U^*_{\alpha 2}U_{\beta 2}]}{\big[(s_1-m_1^2)^2+m^2_1\Gamma_1^2\big]\big[(s_1-m_2^2)^2+m_2^2\Gamma_2^2\big]}\notag\\
			&\quad\quad\times\big[(m_A^2-s_1)(s_1-m_C^2)-s_1s\big]\big[(s_1-m_1^2)(s_1-m_2^2)+m_1m_2\Gamma_1\Gamma_2\big]~.
\end{align}
In contrast, the square terms are
\begin{align}
|\mathcal{M}^{{\rm f}}_j|_+^2 & =\frac{32\ g_+|U^{\alpha j}|^2|U^{\beta j}|^2m_j^2s}{\big[(s_1-m_j^2)^2+m^2_j\Gamma_j^2\big]}~,\\
|\mathcal{M}^{{\rm f}}_j|_-^2 & =\frac{32\ g_-|U^{\alpha j}|^2|U^{\beta j}|^2\big[(m_A^2-s_1)(s_1-m_C^2)-ss_1\big]}{\big[(s_1-m_j^2)^2+m^2_j\Gamma_j^2\big]}~.
\end{align}

In order to obtain the final expression for $d\Gamma^{\alpha\beta}/ds\big|_{{\rm f}}$ we just need to perform the integrations over $s_1$ as in Eq. (\ref{eqn:FDDR}). At leading order in $\epsilon$, the results are displayed in Eqs. (\ref{eqn:IFST}) and (\ref{eqn:IFIT}), but the full coefficients of the interference terms (including terms to all $z$ orders) are 
\begin{align}
	\mathcal{A}^{j}_{+} & = \frac{s}{m^2}\notag\\
	\mathcal{B}^{j}_{+} & = -(-)^j\frac{x}{2}\frac{s}{m^2}(1-z^2)\notag\\
	\mathcal{A}^{j}_{-} & = \mathcal{A}^{j}_{\textrm{sc}} - \frac{s}{m^2}\notag\\
	\mathcal{B}^{j}_{-} & = \mathcal{B}^{j}_{\textrm{sc}} + (-)^j \frac{x}{2} \frac{s m_j^2}{m^4}~.
\end{align}
with $\mathcal{A}^j_{\rm sc}$ and $\mathcal{A}^j_{\rm sc}$ defined in Eq. (\ref{eqn:ASAB}). Note that in the vectorial coupling case, the linear $s$ dependence of the log coefficient terms cancel up to terms of order
\begin{equation}
	\frac{xz}{x^2 + 1} \sim \epsilon \frac{x^2}{x^2 + 1} \le \epsilon~,
\end{equation}
which are negligible, as expected.



%

\end{document}